\documentclass[preprint]{aastex}
\usepackage{emulateapj5}

\shorttitle{Photoevaporation of Circumstellar Disks}
\shortauthors{Font et al.} 

\submitted{To appear in the Astrophysical Journal (received 06/10/03,
accepted 11/02/04)}

\begin{document} 

\title{Photoevaporation of Circumstellar Disks around Young Stars}

\author{Andreea S. Font$^{1}$, Ian G. McCarthy$^{1}$, Doug Johnstone$^{2,1}$, 
and David R. Ballantyne$^{3}$}

\affil{$^1$Department of Physics \& Astronomy, University of Victoria, Victoria, BC, 
V8P 1A1, Canada; afont@uvastro.phys.uvic.ca, mccarthy@uvastro.phys.uvic.ca}

\affil{$^2$National Research Council of Canada, Herzberg Institute of 
Astrophysics, 5071 West Saanich Road, Victoria, BC, V9E 2E7, Canada; 
doug.johnstone@nrc-crnc.gc.ca}

\affil{$^3$Canadian Institute for Theoretical Astrophysics, 60 St George Street, 
Toronto, Ontario, M5S 3H8, Canada; ballantyne@cita.utoronto.ca}

\begin{abstract} 

We examine the ability of photoevaporative disk winds to explain the low-velocity 
components observed in the forbidden line spectra of low-mass T Tauri stars.  
Using the analytic model of Shu, Johnstone, \& Hollenbach (1993) and Hollenbach 
et al. (1994) as a basis, we examine the characteristics of photoevaporative 
outflows with hydrodynamic simulations.  General results from the simulations 
agree well with the analytic predictions, although some small differences are 
present. Most importantly, the flow of material from the disk surface develops 
at smaller radii than in the analytic approximations and the flow-velocity from 
the disk surface is only one-third the sound speed.  A detailed presentation of 
observational consequences of the model is given, including predicted line 
widths, blue-shifts, and integrated luminosities of observable sulfur and 
nitrogen emission lines.  We demonstrate that these predictions are in 
agreement with current observational data on the low-velocity forbidden line 
emission of ionized species from T Tauri stars.  This is in contrast with 
magnetic wind models, which systematically under-predict these forbidden line 
luminosities.  However, the present model cannot easily account for the 
luminosities of neutral oxygen lines in T Tauri stars.

\end{abstract}

\keywords{accretion, accretion disks --- hydrodynamics --- stars: formation}

\section{Introduction}

It is now generally accepted that most young low-mass stars are born with 
circumstellar disks and that the evolution of these disks holds a key to planet 
formation.  Observational evidence for circumstellar disks takes many forms, 
from direct 
imaging (e.g., Jayawardhana et al. 2002) and imaging of disk silhouettes 
(e.g., Bally, O'Dell, \& McCaughrean 2000) through indirect measurements such 
as infrared excess (e.g., Lada et al. 2000). These data provide an estimate for 
the lifetime of disks around low-mass stars, $\tau_{\rm disk} \sim 6 \times 
10^6\,$yr (Haisch, Lada, \& Lada 2001), requiring the presence of an efficient 
disk dispersal mechanism. 

Hollenbach, Yorke, \& Johnstone (2000) considered a variety of disk dispersal
mechanisms and concluded that viscous accretion of the disk onto the central 
star (e.g. Hartmann et al. 1998), although dominating erosion of the inner 
disk, is incapable of removing the entire disk mass in the required time. 
Alternatively, photoevaporation of the disk, while providing an effective 
mechanism for disk removal at large radii $\gtrsim  10\,$AU, cannot remove 
the inner material [Shu, Johnstone, \& Hollenbach 1993 (hereafter SJH93);  
Hollenbach et al. 1994 (hereafter, HJLS94); Johnstone, Hollenbach, \& Bally 
1998].  Acting together viscous accretion and photoevaporation were predicted 
to efficiently remove the entire disk. Numerical calculations by Clarke, 
Gendrin, \& Sotomayor (2001) and Matsuyama, Johnstone, \& Hartmann (2003a) 
have shown that the combined effects of photoevaporation and viscous accretion 
are capable of dispersing disks on $10^5 - 10^7\,$yr timescales.  Additionally, 
the formation of gaps within the gaseous disk during the dispersal era may 
place constraints on the evolution of planetary orbits (Matsuyama, Johnstone, 
\& Murray 2003b).

The photoevaporation model for disk dispersal has been shown to fit the 
observational data well in the case of external heating via nearby massive 
stars (Bally et al. 1998; Johnstone, Hollenbach, \& Bally 1998; St\"orzer \& 
Hollenbach 1998).  However, with the exception of a few cases (e.g., 
MWC349A), evidence for disk photoevaporation due to heating from the central 
star (SJH93, HJLS94) is largely circumstantial.  Low-velocity $\sim 
10\,$km$\,$s$^{-1}$, 
blue-shifted forbidden lines observed in the spectra of T Tauri stars [Hartigan, 
Edwards, \& Ghandour 1995  (hereafter HEG95)] may provide evidence for the 
$10^4\,$K thermal disk winds expected in the photoevaporation model. The 
theoretical models of SJH93 and HJLS94 do not, however, include detailed 
hydrodynamic calculations and thus it has not been possible to test the model 
directly against these observations.

In this paper we take a first look at the hydrodynamic flow of thermally-driven
disk winds powered by photoevaporative heating in order to determine the 
observational consequences of the photoevaporation model. We begin in \S 2 with 
a brief description of how photoevaporation works, concentrating on 
photoevaporation due to heating from a star at the center of the disk. 
In \S 3 we describe how the hydrodynamical simulations are performed and their 
results.  The predicted observable properties of the simulations are 
presented in \S 4 and are compared with existing data.  Further discussion 
and conclusions are presented in \S 5.

\section{Photoevaporation of Circumstellar Disks}

Photoevaporation of disks by ionizing photons is a conceptually simple process to 
understand: high energy radiation (e.g., EUV, X-rays) from the 
central star ionizes hydrogen at the disk surface producing a hot $10^4\,$K 
ionized layer, similar to HII region conditions.  Near the star this coronal zone 
above the disk is expected to be almost static and bound to the disk due to its 
location deep in the potential well of the star, however, at large radii from the 
star the thermal layer should be unbound, powering a thermally-driven disk wind. 
The critical length scale for the model is given in SJH93 and HJLS94 by
\begin{equation}
r_g = { G\,M_* \over c_s^2},
\end{equation}
where $M_*$ is the mass of the star and $c_s \sim 10\,$km\,s$^{-1}$ is the 
sound speed of the ionized gas.

In order to determine the number density of hydrogen atoms at the base of the 
ionized layer SJH93 and HJLS94 assume that for $r < r_g$ the atmosphere is 
in hydrostatic equilibrium such that 
\begin{equation}
n(r,z) = n_0(r)\,\exp\left[-z^2/2\,H(r)^2\right], \ \ \ \ \ (r < r_g)
\end{equation}
where $z$ is the height above the disk and $H(r)$ is the scale height;
\begin{equation}
H(r) = r_g\left(r \over r_g \right)^{3/2}, \ \ \ \ \ (r < r_g).
\end{equation}

The ionization of the material above the disk is not dominated by the direct 
passage of EUV photons from the central star but rather through the 
indirect EUV photons produced by recombination within the ionized layer.  This 
effective 
scattering is strong because one in three recombinations goes directly to the 
ground state, producing a new Lyman continuum photon.  The angular scale 
height, as seen from the central star, grows with distance from the disk 
center enabling the top of the ionized atmosphere to intercept direct EUV 
photons and redirect these photons through recombinations downwards (and 
upwards).  Analytic analysis and numerical computation (HJLS94) allow 
for a determination of the base density power-law profile:
 
\begin{equation}
n_0(r) = n_g\,\left(r \over r_g \right)^{-3/2},\ \ \ \ \ (r < r_g),
\end{equation}
where 
\begin{equation}
n_g = C\,\left( 3\,\Phi_* \over 4\,\pi\,\alpha_2\,r_g^3\right)^{1/2}.
\end{equation}
In the above $\alpha_2$ is the recombination coefficient to all states except 
the ground state and $\Phi_*$ is the ionizing (EUV) photon flux from the 
central star.  The numerical analysis fixes the order unity constant to be $C 
= 0.1$.

At radii larger than $r_g$ the ionized layer is assumed to flow at 
approximately the sound speed and thus the scale height takes the form
\begin{equation}
H(r) = r, \ \ \ \ \ (r > r_g).
\end{equation}
In this case the angular scale height is constant and the atmosphere directly
above the disk no longer has a clear line of sight to the central star; 
therefore, the density at the base of the atmosphere decreases much quicker
than within $r_g$;
\begin{equation}
n_0(r) = n_g\,\left(r \over r_g \right)^{-5/2},\ \ \ \ \ (r > r_g).
\end{equation}
The mass-loss rate due to the flow from both sides of the disk, as a function 
of disk radius, can then be computed directly:
\begin{equation}
{d\dot M_{dw} \over dr} = 4\pi\,r\,n_0(r)\,c_s\,m_{II},\ \ \ \ \ (r > r_g),
\end{equation}
where $m_{II} = 1.35\,m_H$ is the mean mass per hydrogen atom.

Only two parameters are required to fully specify the photoevaporation model: 
$M_*$ and $\Phi_*$.  HJLS94 examined the effects of photoevaporation of the 
circumstellar disk around massive stars where the ionizing flux is set 
by the mass of the star $\Phi_*(M_*)$. SJH93, however, considered 
the effects of 
photoevaporation of the circumstellar disk around low-mass stars and concluded
that only during the early T Tauri stage when the star disk accretion shock
produces enhanced EUV radiation would photoevaporation be important.
Matsuyama et al. (2003a) self-consistently computed the strength of the star disk 
accretion shock during the dispersal of the disk and concluded that high 
ionization rates, $\Phi_* > 10^{40}\,$s$^{-1}$ are expected for the initial $\sim 
10^6\,$yrs. Recently doubts have been raised about the possibility of ionizing
radiation penetrating the thick protostellar jet, which lies between the star
and the disk (Shang et al. 2002). Additionally, the accretion shock itself may
be thick to ionizing radiation (Alexander, Clarke, \& Pringle 2004). In this
paper we assume that a strong ionizing radiation field is present, either from 
the accretion shock, the chromospherically active star, or some other radiation
generating process associated with accretion and the release of potential energy.

A modified photoevaporation model for disk dispersal is applicable to
situations in which the low-mass stars with disks are located near massive
stars, such as in the Orion Trapezium.  Johnstone et al. (1998) showed that in
such cases the geometry for both the radiative transfer of the ionizing radiation
and the hydrodynamic flow is much simpler and the model results provide a good 
match to the observations.  However, it is clear that such external 
evaporation is not applicable to every low-mass star.  In the case of 
photoevaporation due to an isolated central star, detailed fits to the 
observational data have not been possible due to the poor understanding of the 
flow characteristics.  Yet there are clear strengths to such a model, including
the ability to disperse the disk on the observationally determined timescale 
(SJH93; Hollenbach et al. 2000; Matsuyama et al 2003a) and the ability to 
produce a thermally-driven wind with characteristic speed $c_s$ similar to 
low-velocity, blue-shifted forbidden 
line value observed in T Tauri star spectra (HEG95). The photoevaporation model 
also requires only a few free parameters and thus the observational characteristics 
should be well determined.  This aspect of the model is in direct contrast with 
the magnetic wind calculations often used to explain the low-velocity, 
blue-shifted forbidden line radiation (Cabrit, Ferreira, \& Raga 1999; Garcia
et al. 2001a, 2001b).

\section{Hydrodynamic Disk Wind Models}

As discussed in \S 2, a detailed and complete picture of the thermal disk wind 
can only be attained by explicitly solving for the hydrodynamic flow 
characteristics.  For this purpose, we make use of the publicly available 
ZEUS-2D hydrodynamical code (Stone \& Norman 1992a, 1992b; Stone, Mihalas, \& 
Norman 1992).  This code has been used in previous studies of disk outflows 
(e.g., Lee et al. 2001) and is a good choice because of its great flexibility in 
terms of geometry and grid spacing.

The effects of magnetic fields, radiation, radiative transfer, dust, and 
self-gravity of the gas are neglected in these models.  We also neglect the 
effects of a stellar wind or an inner jet on the dynamics of the flow.  
A discussion of how this omission could potentially influence our results and 
conclusions is provided in \S 5.

\subsection{A spherical test: the Parker wind problem}

Before moving on to the problem of winds from photoevaporating circumstellar 
disks, it is instructive to consider a simpler problem to test the method.  
The Parker solar wind problem (Parker 1963) is ideal for this purpose because 
it is a trans-sonic outflow problem with a known solution.  The Parker wind 
problem considers the outflow properties of a thermally-driven spherically symmetric 
wind from a gravitating point-mass (such as the Sun).  Assuming the wind is 
isothermal and barotropic (i.e., the pressure exerted on the gas is a function of 
density only), then there is a well-defined analytic solution with the wind 
passing through a sonic point ($|v|/c_s = 1$) at the radius $r/r_g = 1/2$.  

To set-up the problem in ZEUS-2D, a simple spherical coordinate grid is utilized.  
The radial coordinate, which is chosen to span $r/r_g = 0.1 - 1 .0$, is divided 
linearly into 100 cells, while the polar coordinate, for which the range $\theta 
= 0^{\circ} - 90^{\circ}$ is used, is divided linearly into 60 cells.  The flow 
is assumed to be azimuthally symmetric in ZEUS-2D.  The boundary conditions of 
the simulation are all made to be reflecting except the outer radial boundary, 
which is set as an outflow boundary, and the inner boundary, which is 
continually updated such that the density remains constant along the inner edge 
(representing the source of the wind, i.e., near the surface of the star).
The density at this inner edge was normalized to 1 while the remainder of
the cells in the grid were set to an initial density of $10^{-4}$.
The initial velocity within each cell is set to zero and the simulation is 
allowed to run until it reaches steady-state.  In practice, the velocity of 
the gas as a function of radius and time is tracked and we verify that the 
simulation converges rapidly to a steady-state solution identical (to within a 
few percent accuracy) to the analytic solution, with the sonic point occurring 
at $r/r_g = 1/2$.  Since the simulation setup just presented is 
very similar to the required setup for the disk wind models discussed below, 
the Parker wind analysis provides a good test of their reliability.

\subsection{A disk wind test: power-law models}

Below we explore winds originating from simple power-law disk models.  
Before proceeding, we emphasize here that these models are of academic interest 
only.  Our purpose is to examine and understand the character of their solutions.  
This, in turn, provides a further test of our methods and also gives a 
basis for understanding the more realistic hybrid case considered below.

\subsubsection{Setup}

We now consider winds from photoevaporating disks.  First, a description is 
given of the simulation setup in ZEUS-2D and how the models are initialized.

At large radii the flow is expected to be dominated by pressure gradients in 
the diverging flow and should asymptotically become radial.  Therefore, a 
spherical polar coordinate system is selected for these simulations.  Since it 
is not known {\it a priori} where the flow will become sonic, the radial 
coordinate spans a large range, $r/r_g = 0.1 - 20$ and is divided 
logarithmically into 200 cells.  The polar coordinate is chosen to span the 
range $\theta = 0^{\circ} - 90^{\circ}$ (thus, we are simulating just one 
quadrant of the atmosphere) and is divided linearly into 100 cells. By using 
this setup, we avoid excessive computation times but still have high enough 
resolution near the base, where the pressure gradients are largest, to 
accurately capture the dynamics of the flow.

Similar to the Parker wind model described above, all of the boundaries are made 
to be reflecting except the outer radial boundary, which is an outflow boundary, 
and the disk base, which is updated after each time step to the original 
density distribution (representing replenishment from the photoevaporation of 
the neutral disk).  The disk directly below the ionized atmosphere is assumed to 
be an infinitely flat, large (i.e., occupies the entire base of the grid) 
reservoir of neutral material.  The base density in the ionized atmosphere is 
initially set-up as a simple power-law distribution with
\begin{equation}
n_0(r) = n_g\,\biggl(\frac{r}{r_g}\biggr)^{-\alpha},
\end{equation}
where $\alpha = 3/2$ or $5/2$ is the power-law index, spanning the range 
of model conditions (i.e. Eqns. 4 and 7).

The ionized disk surface is laid down on the equatorial axis ($\theta = 90^{\circ}$) 
and the density of the disk (and the above atmosphere) is normalized such that 
$n_g = 1$.  The disk is assigned no velocity except for a Keplerian angular 
velocity.  As done for the Parker wind problem, the remaining cells, which 
comprise the `atmosphere', are initially filled with a low-density 
($10^{-4}\, n_g$) gas and are assigned no velocity. The results of the 
simulation are not at all sensitive to the exact value of this density.  The 
simulations are allowed to run until they reach steady-state solutions.  

\subsubsection{Disk wind test results}

Plotted in Figures 1a and 2a are steady-state streamlines and velocity contours 
for the two test cases $\alpha = 3/2$ and $\alpha = 5/2$.  The streamlines 
(solid lines) are lines of constant angular momentum, $j_{\phi}$, (which is a 
conserved quantity) and demonstrate how the ionized gas flows from the disk.  
They are spaced in Figs. 1 and 2 according to the total disk wind mass-loss 
rate, $\dot{M}_{dw,tot}$ (see the caption of Fig. 1).  A comparison of the two 
figures clearly shows that a much larger fraction of the total mass loss comes 
from the central regions of the $\alpha = 5/2$ case than of the $\alpha = 3/2$ 
case.  Furthermore, the streamlines in the $\alpha = 5/2$ case are approximately 
radial, whereas the streamlines in the $\alpha = 3/2$ case contain a more 
significant vertical component.

The differences in the spacings and directions of the streamlines of the two 
cases can be understood physically.  Since the $\alpha = 5/2$ case has a 
steeper base density profile than the $\alpha = 3/2$ case, a larger fraction 
of the ionized gas mass is located within the central region of that model.  
{\epsscale{1.0}
\plotone{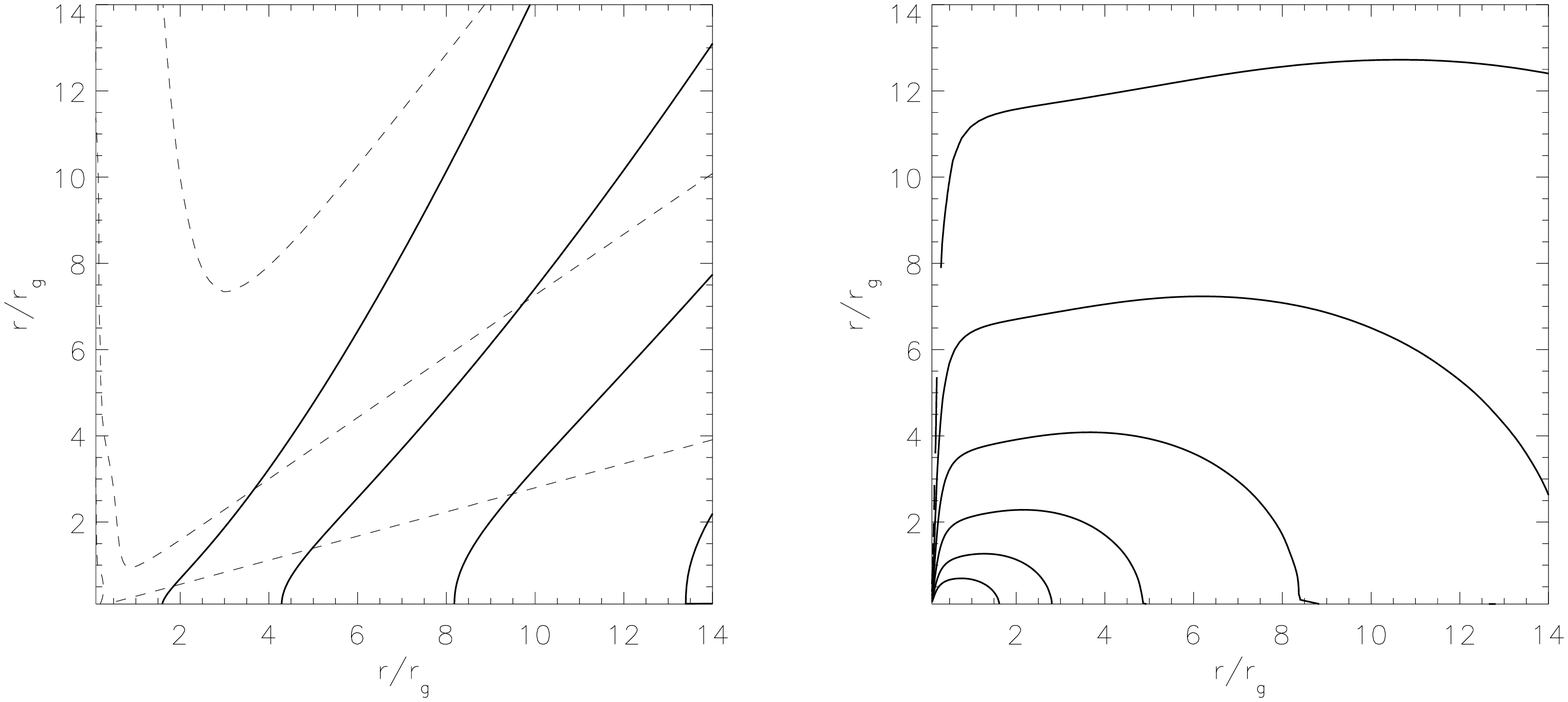}
{Fig. 1. \footnotesize
Steady-state results for the $\alpha = 3/2$ case.  {\it Left}: Streamline and
velocity contours.  The solid lines are streamlines and indicate how the ionized
gas flows off the disk.  They are spaced evenly according to the total mass-loss
rate.  Namely, the streamlines enclose 20\%, 40\%, 60\%, 80\%, and 95\% (not
shown) of $\dot{M}_{dw,tot}$.  The dashed lines indicate where $v_{tot}$ equals
$1$, $2$, and $3$ times the sound speed (with the velocity of the flow
increasing with distance from the disk).  {\it Right}: Density contours.
Contours are evenly separated in log-space and range from $\log{n/n_g} = -2.5$
to $0$.
}}
\vskip0.1in
{\epsscale{1.0}
\plotone{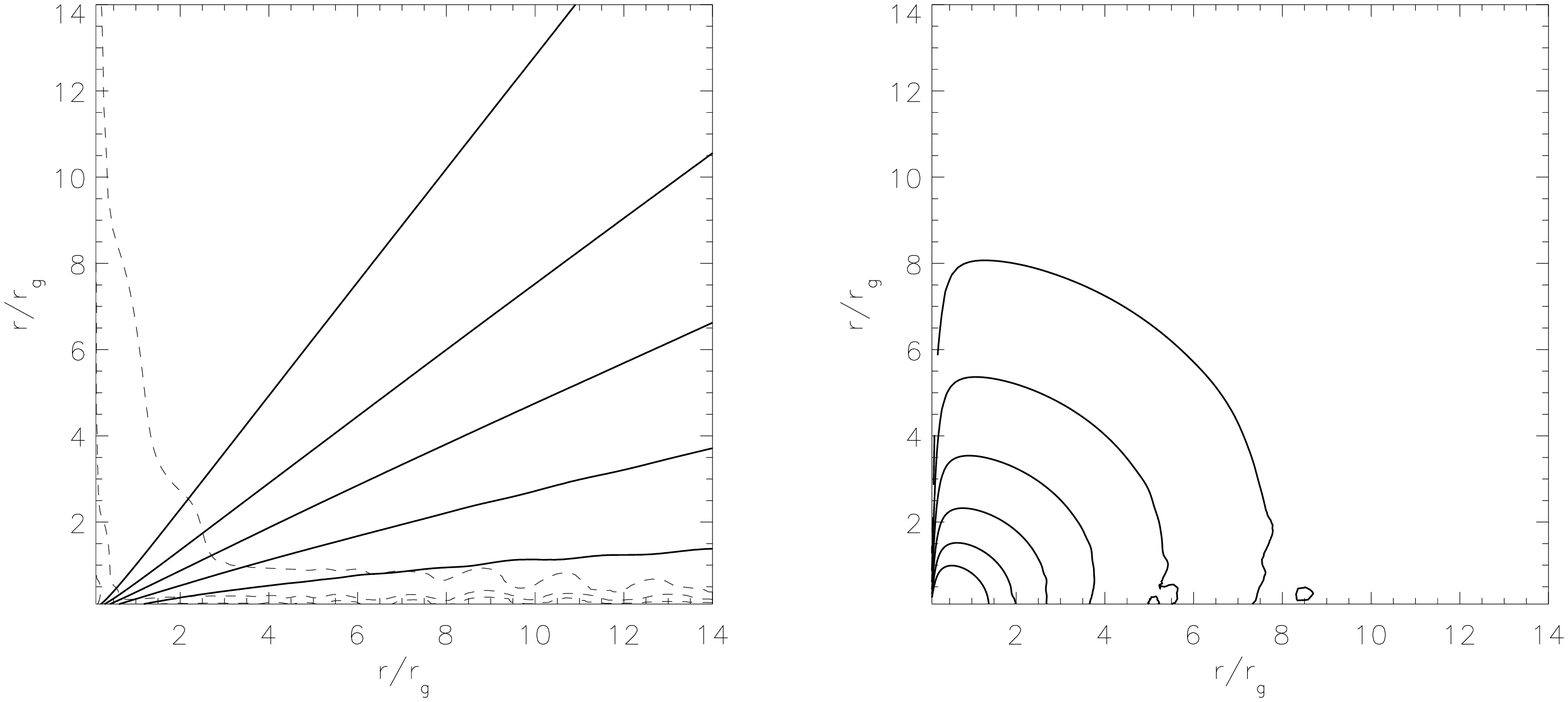}
{Fig. 2. \footnotesize
Steady-state results for the $\alpha = 5/2$ case.  The panels and lines have the
same meaning as in Fig. 1.
}}
\vskip0.1in
\noindent
Indeed $\sim$ 90\% the total mass loss comes from inside $r_g$ for the $\alpha = 
5/2$ case.  In addition, because the $\alpha = 5/2$ case has a steeper density 
profile, this necessarily implies that it has a steeper pressure gradient as 
well.  The steeper pressure gradient forces the streamlines to bend more quickly 
with respect to the vertical direction, producing a more spherical outflow.

The dashed lines in Figs. 1a and 2a represent total velocity ($v_{tot} = 
\sqrt{v_{\theta}^2 + v_{r}^2}$) contours.  The lines demonstrate where 
$v_{tot}$ is equal to $c_s$, 2 $c_s$, and 3 $c_s$.  It is clear that the 
gas flows off the disk in the $\alpha = 5/2$ case much faster than in the 
$\alpha = 3/2$ case, directly related to the pressure gradients acting on the
system.  
  
Intimately linked with the velocity and streamline properties is the atmosphere 
density distribution.  Plotted in Figs 1b and 2b are density contours for the two 
cases.  The contours are evenly spaced in $\log{}$-space and range from 
$\log{n/n_g} =$ -2.5 to 0.  The $\alpha = 3/2$ case shows horizontally elongated 
contours while the $\alpha = 5/2$ case shows nearly spherical contours.  Again, 
the difference arises because the $\alpha = 3/2$ case has an appreciable mass-loss 
rate even at large radii, while $\alpha = 5/2$ case has almost no mass loss at 
large radii.
  
The small irregularities visible at large radii (near the base) in the
$\alpha = 5/2$ case are transient phenomena (not steady-state) and are due to the 
fact that the density of the base at these radii is comparable to that of the 
outflowing atmosphere and, therefore, the flow is easily disrupted.  The low 
densities and mass-loss rates at these radii, 
{\epsscale{1.0}
\plotone{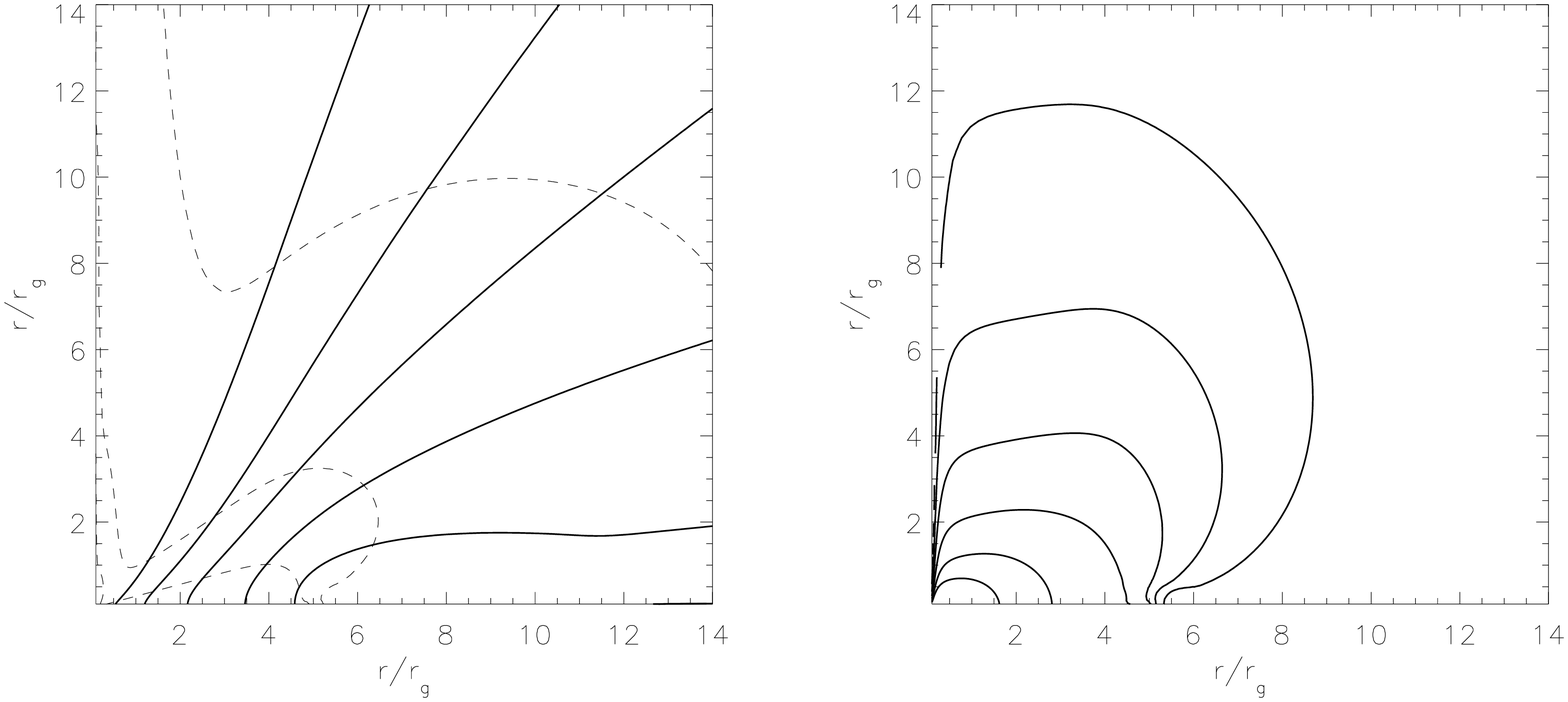}
{Fig. 3. \footnotesize
Steady-state results for the truncated $\alpha = 3/2$ case.  The ionized disk
has been truncated at $r/r_g = 5$.  The panels and lines
have the same meaning as in Fig. 1.
}}
\vskip0.1in
\noindent
however, ensure that these 
irregularities have little or no effect on the rest of the (steady-state) 
atmosphere or on the predicted observable properties of these models.

\subsubsection{A truncated disk}

Figure 1 illustrates that a significant fraction of the total mass-loss 
rate for the $\alpha = 3/2$ case originates at large radii.  This gives rise to 
the elongated density contours shown in Fig. 1b.  Furthermore, because of the 
shallow pressure profile of the ionized disk wind, the streamlines show large 
vertical components (i.e., perpendicular to the disk).  We test these ideas 
quantitatively by running another $\alpha = 3/2$ simulation but this time with 
the disk truncated at $r/r_g = 5$.  If the reasoning above is correct, we 
should expect the density contours for the truncated case to be more spherically 
symmetric while streamlines originating near the truncation radius should 
show greatly reduced vertical components (as there is no force at large 
radii pushing the gas vertically).

Plotted in Figure 3 are the streamlines and velocity and density contours for 
the truncated $\alpha = 3/2$ case.  As expected, the streamlines (Fig. 3a) 
illustrate that the flow is becoming approximately radial at large radii.  Of 
note is the streamline which encloses 95\% of the total mass-loss rate (and 
originates quite close to the truncation radius), which is essentially 
parallel to the ionized disk.  Moreover, the density contours (Fig. 3b) 
also show a more spherical outflow than for the original $\alpha = 3/2$ 
case plotted in Fig. 1b.  At large radii, the flow tends to spherical 
symmetry.

\subsubsection{Comparison to the Parker wind}

In the $\alpha = 3/2$ case the disk mass loss is dominated by loss from
the disk at large distances from the central star, producing a non-spherically 
symmetric outflow even at large radii.  Alternatively, in the $\alpha = 5/2$ 
case the disk mass loss is dominated by the central region and the far field 
limit tends to spherical symmetry.  Comparison of the evaporating disk results 
with the analytic Parker wind solution provides an ideal framework for 
understanding the hydrodynamic flow.

In Figure 4, the velocity and density profiles of the two cases are compared 
to the analytic solution to the Parker wind problem (assuming isothermal, 
barotropic gas).  The simulation profiles are derived at a fixed polar coordinate 
$\theta = 45^{\circ}$ but could have just as easily been chosen from any other 
slice through the atmosphere (except near the extremes $\theta \approx 0^{\circ}$
or $90^{\circ}$).

The results plotted in Fig. 4a clearly indicate that the 
{\epsscale{1.0}
\plotone{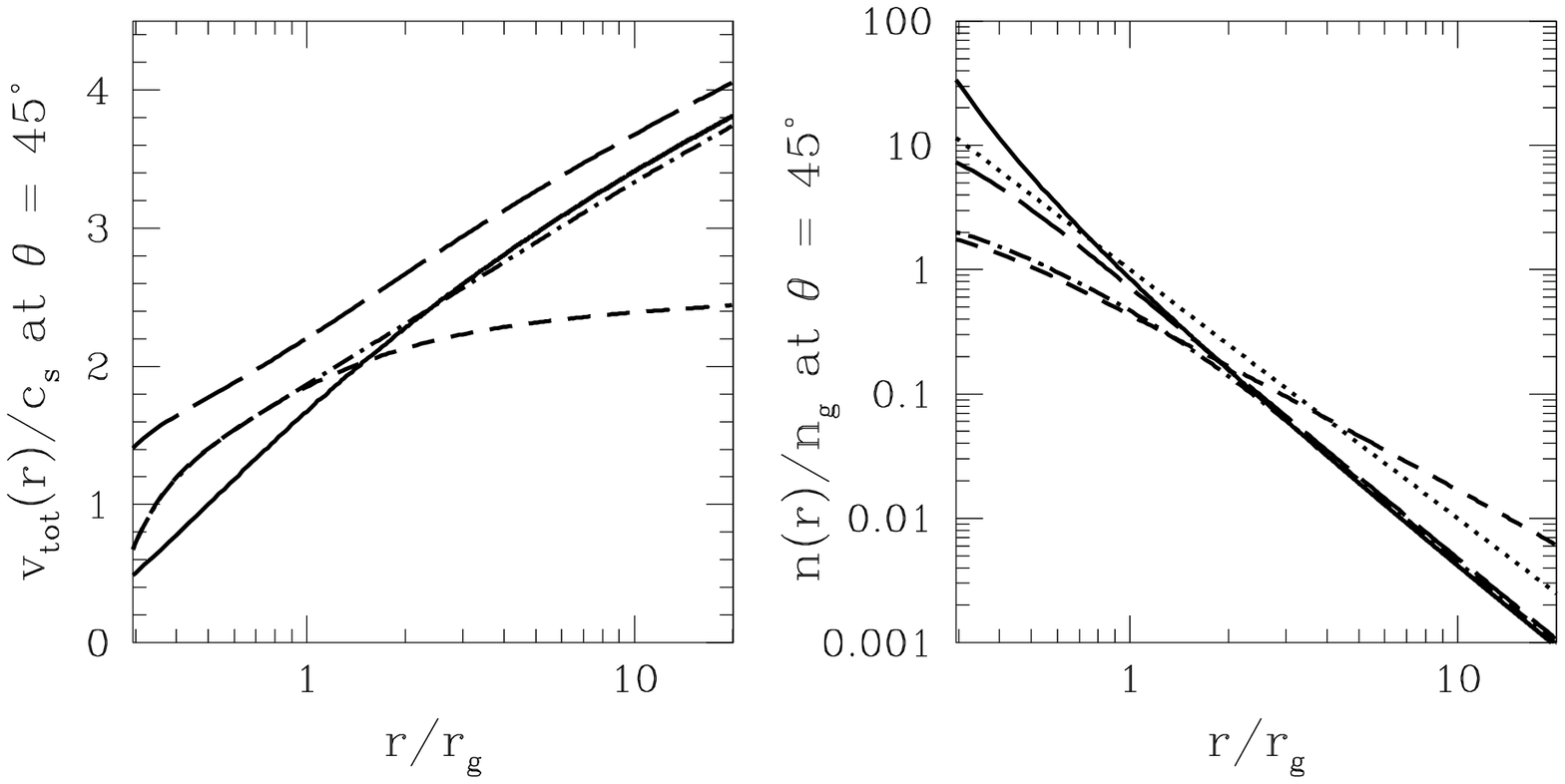}
{Fig. 4. \footnotesize
Velocity and density profiles at $\theta = 45^{\circ}$.  {\it Left}: Velocity
profiles.  {\it Right}: Density profiles.  The short dashed line represents the
$\alpha = 3/2$ case, the long dashed line represents the $\alpha = 5/2$
case, and
the dot dashed line represents the PDW model (see \S3.3).  Also shown
(solid line) is the prediction of the Parker wind model.  The density profile
of the Parker wind model has been arbitrarily normalized to match the other
cases near $r/r_g \approx 2$.  For comparison, the dotted line in the right
hand panel shows $n \propto r^{-2}$.
}}
\vskip0.1in
\noindent
shape of the velocity 
profile for the $\alpha = 5/2$ case closely matches that of the 
Parker wind solution, including the normalization.  The $\alpha = 3/2$ 
case, however, has a much different shape and normalization than the 
Parker wind profile.  This simply reflects the fact the $\alpha = 3/2$ has a 
significant mass-loss rate even at large radii and the pressure forces from 
this gas are still modifying the dynamics of the flow (i.e., it is not yet 
dominated by spherical divergence).

Fig. 4b illustrates that the flow from the $\alpha = 5/2$ case is 
thermally-driven at large radii.  For $r \gtrsim r_g$, the 
density scales almost exactly as $r^{-2}$, as expected for a thermally-driven 
flow (see the Parker wind line, for example).  The $\alpha = 3/2$ case, 
however, shows an over-dense flow at large radii with respect to the other 
cases.  Again, this may be attributed to the non-negligible mass-loss rate at 
large radii in this case.

\subsubsection{Comparison to the analytic model}

The simulations behave as expected, at least at large radii, and we are now in a 
position to compare the steady-state properties with the model of SJH93 and 
HJLS94 (from which we derive our physically-motivated initial conditions).  
As discussed in \S 2, hydrodynamic simulations, such as those performed here, 
are required for a proper treatment of the flow dynamics.  Thus, the model of 
SJH93 and HJLS94 incorporated a number of simplifying assumptions about the 
form of the atmosphere, especially for the scale height $H(r)$.  In this 
section we test whether these assumptions are valid or not.

In the SJH93 and HJLS94 model, the density distribution within $r_g$, their 
``no flow'' region, is determined entirely by the base density distribution, 
$n_0(r)$, and by the scale height of the ionized atmosphere, $H(r)$.  Since we 
assume a base density distribution that is nearly identical to theirs, any 
differences between the models will manifest themselves in differences between 
the scale height distributions.  Plotted in Figure 5 are the predicted scale 
height profiles for the $\alpha = 3/2$ and $\alpha = 5/2$ cases.  Also shown is 
the analytic result of SJH93 and HJLS94.  Reassuringly, each of the cases 
closely follow the analytic result out to $r/r_g \sim 0.4$, matching both the 
shape and normalization of the analytic result.  They each break away from the 
{\epsscale{1.0}
\plotone{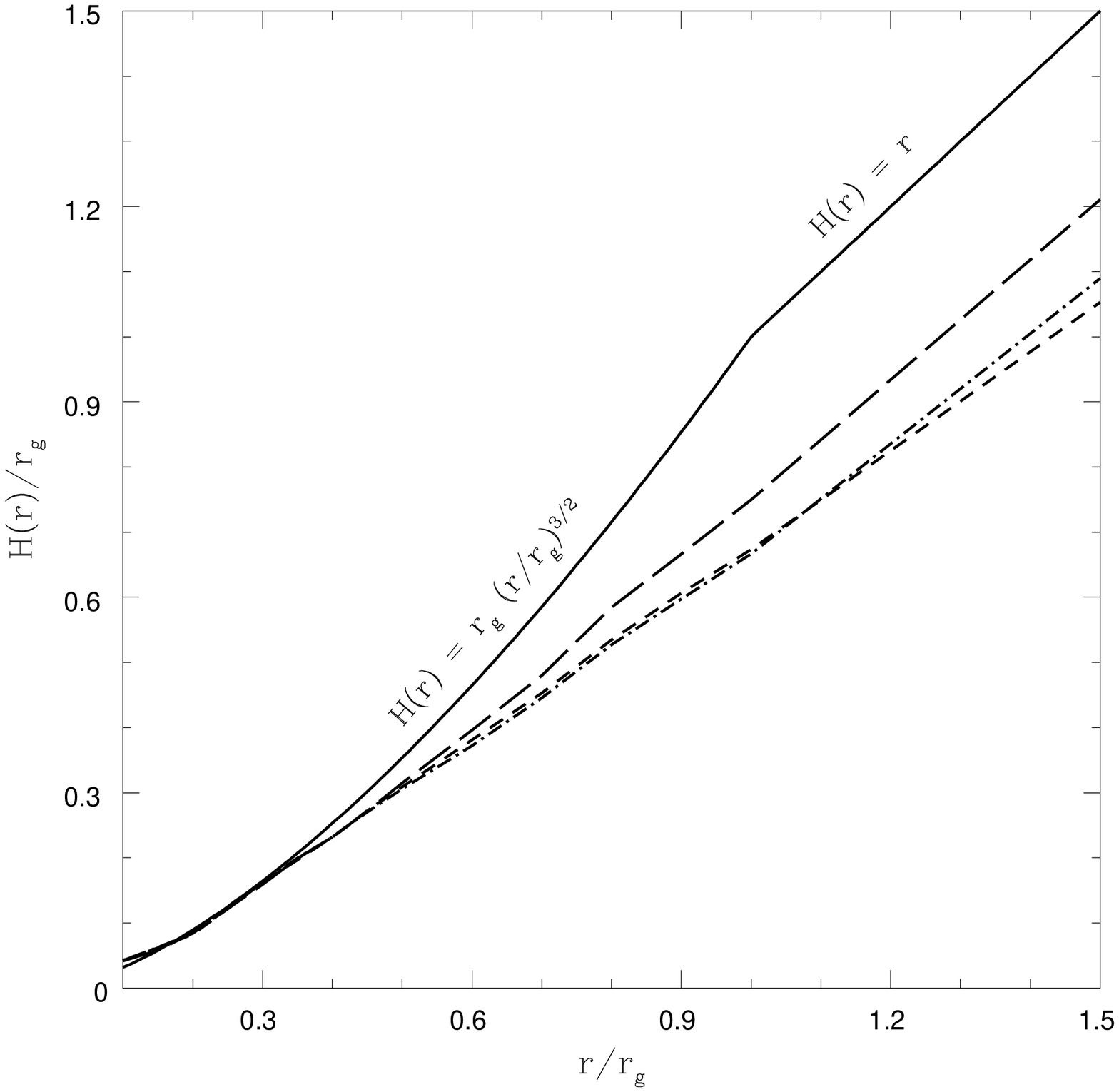}
{Fig. 5. \footnotesize
The scale height of the ionized atmosphere as a function of disk radius.  The
solid line is the analytic result of SJH93 and HJLS94.  The short dashed line
represents the $\alpha = 3/2$ case, the long dashed line represents the
$\alpha = 5/2$ case, and the dot dashed line is for the PDW model (see \S3.3).
}}
\vskip0.1in
\noindent
predicted relation at larger radii.  This is not unexpected as the 
identification of $r_g$ as the launching point for the outflow is only accurate 
up to an order unity coefficient.  For example, the Parker wind `launches' at 
$0.5\,r_g$ due to pressure gradients in the divergent flow, and the disk wind 
can produce an even  stronger pressure gradient.  The fact that the simulations 
match the analytic result out to $r/r_g \sim 0.4$ implies that simulations and 
analytic model are neither hugely discrepant, nor are they strongly influenced 
by the base density distribution [at least in determining $H(r)$].  Thus, we 
are confident that a hydrodynamic model with a shallow base density slope near 
the star and a steeper density slope at large radii (such as that considered 
immediately below) will produce a reasonable model of the photoevaporating 
disks.

\subsection{Photoevaporative disk wind (PDW) model}

While the $\alpha = 3/2$ and $\alpha = 5/2$ cases were used to represent the 
range of base conditions, neither provides a reasonable fit to the full range 
of densities at the base of the photoevaporating disk models developed by SJH93 
and HJLS94.  A better solution is to use a hybrid power-law model, which has a 
base density profile that mimics the $\alpha = 3/2$ case at small radii and the
$\alpha = 5/2$ case at large radii.  In fact, the radiative transfer 
calculations of HJLS94 demand such a model.  We choose the following fitting 
function in order to quickly yet smoothly make the transition: 
\begin{eqnarray}
n_0(r)  = n_g \biggl(\frac{2}{[(r/r_g)^{15/2} + (r/r_g)^{25/2}]}\biggr)^{1/5}.
\end{eqnarray}
Alternative fitting functions were tested and no significant changes in the
results below were found.  This hybrid 
{\epsscale{1.0}
\plotone{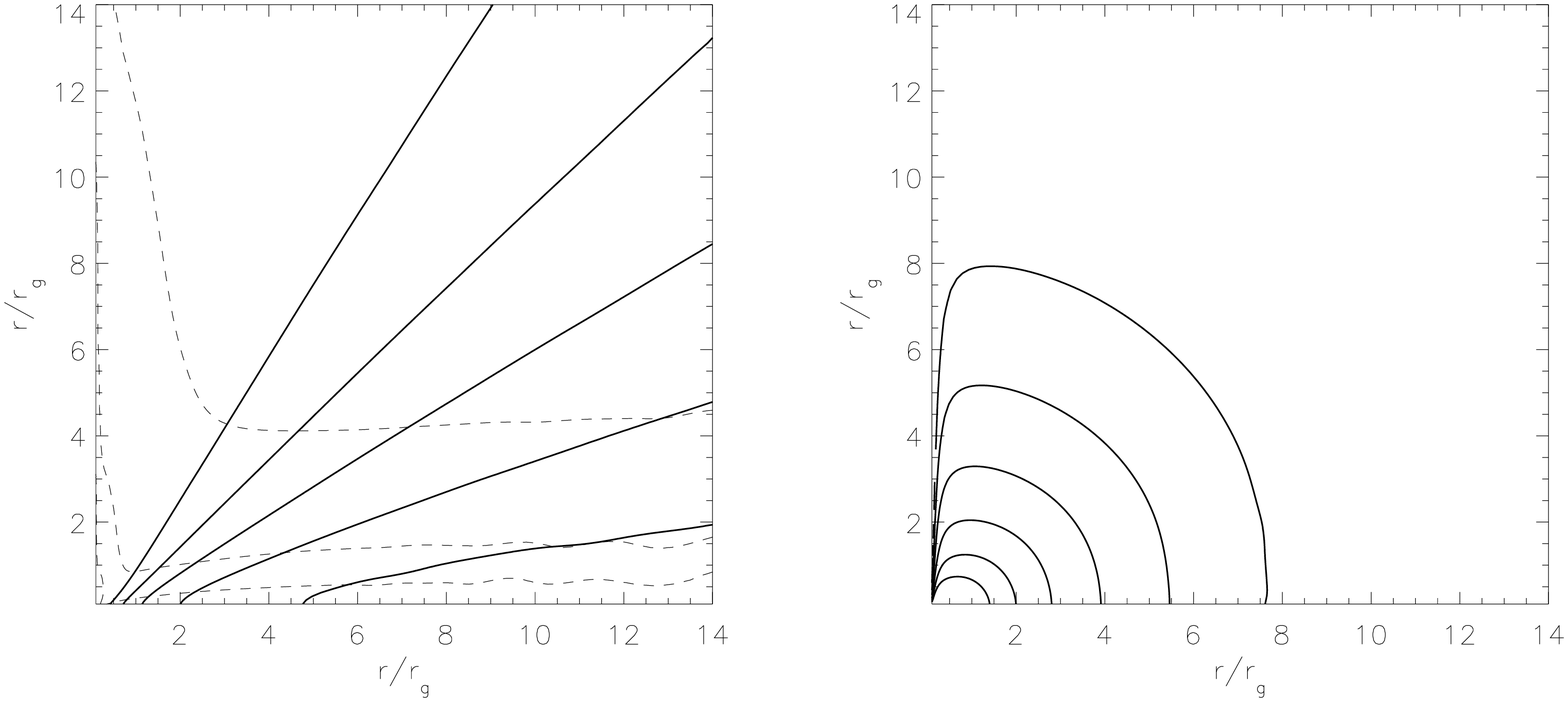}
{Fig. 6. \footnotesize
Steady-state results for the photoevaporative disk wind model.  The panels and
lines have the same meaning as in Fig. 1.
}}
\vskip0.1in
\noindent
model, which, henceforth, will be 
referred to as the photoevaporative disk wind (PDW) model, is run in ZEUS-2D 
under the same conditions as the $\alpha = 3/2$ and $\alpha = 5/2$ cases, as 
described in \S 3.2.1.  

Plotted in Figure 6 are streamlines, velocity, and density contours for the 
PDW  model.  A quick comparison of Fig. 6 with Figs. 1 and 
2 reveals that, as expected, the PDW model is intermediate between the $\alpha = 
3/2$ and $\alpha = 5/2$ cases.  However, it more closely resembles the $\alpha = 
5/2$ case.  The reason for this is that most of the total mass-loss rate in the 
PDW model is confined to $r/r_g \lesssim 2$, which is very similar to the 
$\alpha = 5/2$ case.  In addition, both models appear to be approaching 
spherical symmetry at large radii (the density contours are nearly spherical 
while the streamlines are approximately radial).  This is an expected result.  
Figs. 4 and 5 compare the results for the PDW model flow against the other two 
cases and the Parker wind solution.

A physically insightful comparison between the PDW model and the analytic 
model of SJH93 and  HJLS94 is through the integrated mass-loss rate profile.  
The integrated mass-loss rate along the disk, assuming azimuthal symmetry and 
for both disk hemispheres, is calculated via
\begin{equation}
\dot{M}_{dw}(r) = 4 \pi \int_{0.1r_g}^{r} \rho(r') 
v_{\theta}(r') r' dr' .
\end{equation}
\noindent The mass-loss rate is presented in dimensionless units in Fig. 7.  

There are two noticeable differences between the PDW model and the 
analytic result.  First, the PDW model shows significant mass-loss at radii 
smaller than $r_g$, whereas the analytic model has no mass-loss at these radii (by 
definition).  Also clearly evident is that the PDW model predicts a lower 
total integrated mass-loss rate than the analytic result (by approximately a 
factor of $2.7$).  This is due entirely to the velocity of the flow off the 
disk being much smaller than that assumed by SJH93 and HJLS94 (they assumed the 
flow velocity is equal to the sound speed).  

The total mass-loss rate is determined by both the base velocity {\it and} 
density. Since the models used here fix the density along the base to be 
nearly identical to the analytic result of HJLS94, it is the change in the 
launch velocity which accounts for the difference in the mass-loss rate.  Our 
simulations have outflow velocities in the range of $\sim 0.3-0.4 c_s$.  As a 
check, if we {\it assume} $v_{\theta} = c_s$ and integrate (11) from $r/r_g = 1$ 
to $20$ (to mimic the analytic result), we find a total mass-loss rate that is 
within 10\% of the analytic result. 
 
The velocity profile of the flow is determined primarily 
{\epsscale{1.0}
\plotone{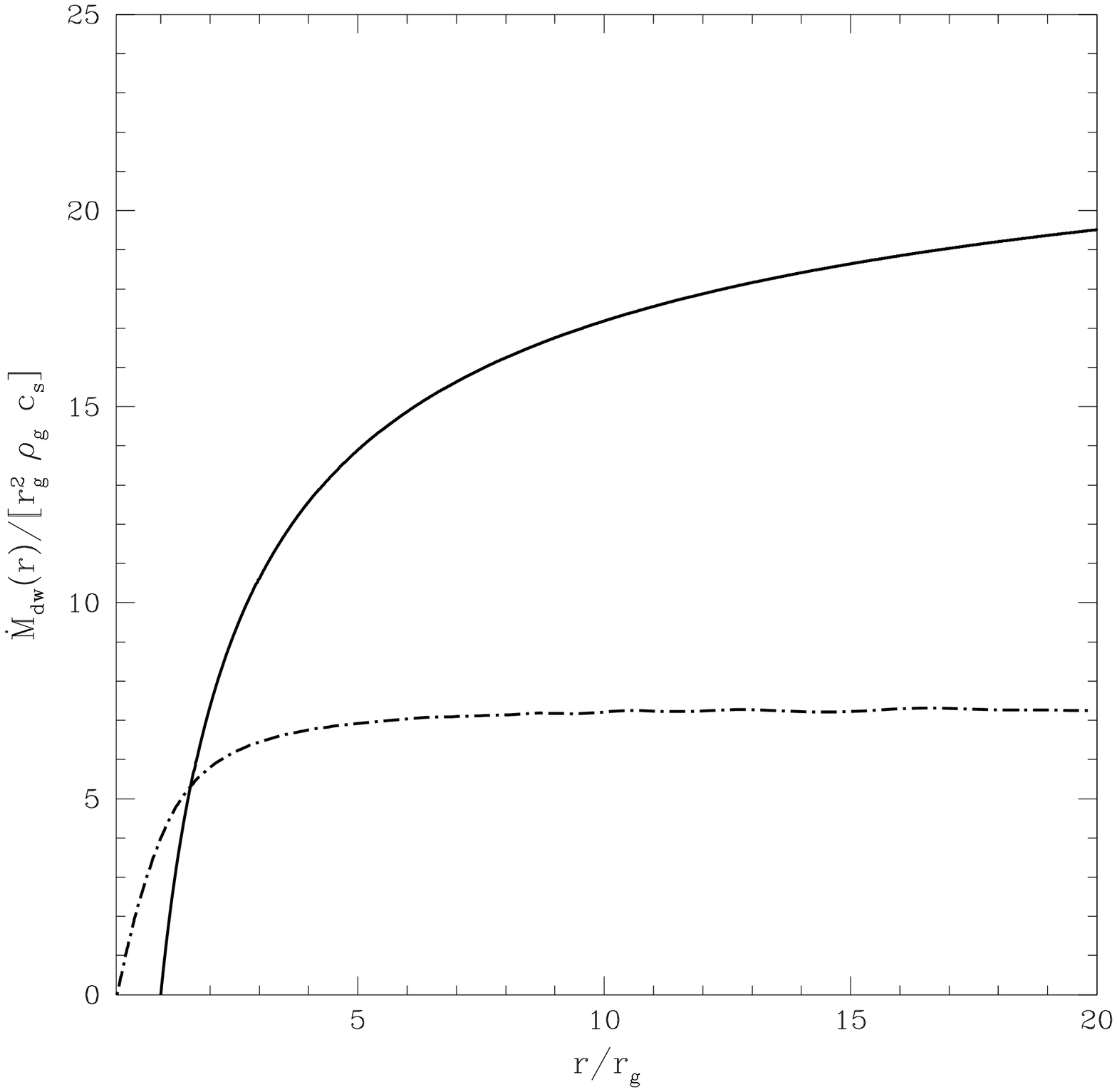}
{Fig. 7. \footnotesize
Integrated disk wind mass-loss rates as a function of disk radius.   The dot
dashed line represents the PDW model, while the solid line represents the
analytic result of SJH93 and HJLS94.
}}
\vskip0.1in
\noindent
from the divergence
of the streamlines and depends on density contrast, not the absolute density.
Provided that the density drops as steeply as $\alpha = 5/2$ in the outer
disk, the base velocity should be reasonably well determined by the
hydrodynamic simulation. The normalization of the density, however, depends
on the detailed balance between incident ionizing radiation, recombinations
within the flow, and the liberation of new ions at the base. HJLS94 solved
this balance equation explicitly, assuming a constant flow, and found that 
near $r_g$, where the flow develops, recombinations dominate or are comparable 
to the production of new ions. Thus the base density varies smoothly through 
$r_g$ with the power-law change reflecting the diminishing importance of overhead 
redirected ionizing radiation. This may need to be modified somewhat for the present
model due to the acceleration of the flow and resultant drop in the number density
of ionized hydrogen. Thus, the mass-loss rate presented here should be thought of
as a lower-limit. Farther from the central star the new ions liberated at the base 
of the disk will eventually dominate the detailed balance, resulting in a 
{\it steeper} density profile and even less evaporation.  The density determinations
of HJLS94 are {\it strengthened} when a lower base velocity is adopted.

What are the implications of our simulations for the study of photoevaporating
disks?  A primary goal of HJLS94 was to see whether the photoevaporation of 
neutral disks could replenish the expanding ionized gas in ultra-compact HII 
regions, thus explaining the paradoxical long lifetimes of such systems.  In 
this context, the lifetime of an HII region is inversely proportional to the 
disk wind mass-loss rate. Since the simulations predict mass-loss rates that 
are within a factor of $2-3$ of the HJLS94 model, our results do not 
significantly modify the conclusions of these authors.  Rather, they reaffirm 
the importance of the photoevaporation mechanism proposed by HJLS94.  The same 
situation holds for the dispersal of disks around low-mass stars (SJH93; Clarke 
et al. 2001; Matsuyama et al. 2003a). We do not expect major changes in the 
conclusions of those papers when these more accurate dispersal values are used.
Only the details of how the disk evolves near $r_g$, where the hydrodynamic 
simulations show the strongest disagreement with the analytic theory, should be 
effected.  Thus, the importance of a gap forming within the disk around $r_g$ 
and preventing planetary migration may need to be recalculated (Matsuyama et al. 
2003b).

After the work of SJH93 and HJLS94 there has been a number of published 
observational studies on low-velocity outflows in young stellar systems.  
Most notable of these is the T Tauri survey of HEG95.  Below, we consider the 
observational consequences of our simulations for photoevaporating disks 
around young, low-mass T Tauri stars.  We then compare our results with the 
observational data of HEG95.  This is the focus of \S 4.  We show that the 
simulations generally give reasonable fits to a number of observed 
trends.

\section{Observational Consequences and Comparison to HEG95}

The most common way to constrain the properties of low-velocity outflows from 
T Tauri stars is through observations of forbidden emission lines (typically, 
optical lines).  Commonly measured lines include [SII]$\lambda6731$, 
[SII]$\lambda6716$,  [NII]$\lambda6583$, [OI]$\lambda6300$, and 
[OI]$\lambda5577$.  Here, we explore the predicted profiles and strengths of 
such lines from our simulations and how these depend on, for example, 
inclination angle of the system and what region of the system is being observed 
(in the event the system is resolved).  We also compare the model results 
with the observational data of HEG95.  

\subsection{Rescaling for T Tauri systems}

Before line profiles and luminosities can be calculated, the simulations must 
be scaled from dimensionless units into physical units.  The three fundamental 
quantities that must be rescaled are the velocity of the gas, the density of the 
gas, and the size of the system.  

First, the velocity of the gas is rescaled by considering that for a gas with a 
solar abundance of elements and a temperature of $10^4$ K, the sound speed, 
$c_s$, is $\approx 10$ km s$^{-1}$.   
  
The physical size of the system is calculated by rescaling $r_g$.  Assuming a 
sound speed of $10$ km s$^{-1}$, the value of $r_g$ is given by
\begin{equation}
r_g \approx 1.3 \times 10^{14} \biggl(\frac{M_*}{1 M_{\odot}}\biggl) 
\, \, \, {\rm cm}
\end{equation}

\noindent In what follows, we focus on systems with $0.1 M_{\odot} \leq M_* 
\leq 2 M_{\odot}$, similar to those observed by HEG95.

Finally, we re-scale the density of the simulations.  As discussed in SJH93 and
HJLS94, the value of $n_g$ is determined by the mass of the star (or, 
equivalently, $r_g$) and by the rate of ionizing photons, $\Phi_*$, that are 
emitted by the star
\begin{equation}
n_g \approx 4 \times 10^4\, \biggl(\frac{M_*}{1 M_{\odot}}\biggr)^{-3/2} 
\biggl(\frac{\Phi_*}{10^{41} s^{-1}} \biggr)^{1/2} \, \,
\, {\rm cm}^{-3}
\end{equation}  

\noindent We assume rates of $10^{40}$ s$^{-1}$ $\leq \Phi_* \leq 10^{42}$ 
s$^{-1}$, which is compatible with what is expected from T Tauri stars that are 
accreting gas on to their surfaces (e.g., Matsuyama et al. 2003a).  In \S5,
we discuss the validity of this assumption.

The forbidden line luminosity, $L$, at a velocity $u$ is given by
\begin{equation} 
L(u) = \frac{1}{\sqrt{2 \pi c_s^2}}\int e^{-\frac{(u - \mu)^2}{2 
c_s^2}} \epsilon(n,T) n(\theta,r) \frac{n_s}{n} dV  \, \, ,
\end{equation} 

\noindent where $\mu$ is the velocity shift along the line-of-sight due to the 
motion of the gas, $\epsilon$ is the line emissivity per ion, $n_s/n$ is 
abundance of the atomic species from which the line originates, and $dV = r^2 
\sin{\theta} d\theta d\phi dr$.  We have assumed that each cell in the simulation 
emits a Gaussian line profile, which is re-normalized by $(2 \pi c_s^2)^{-1/2}$ 
to give unity when integrated.

The velocity shift, $\mu$, due to a particular cell's motion is given by
\begin{eqnarray}
\mu= [(v_{\theta} \cos{\theta} + v_r \sin{\theta}) \cos{\phi} - 
v_{\phi} \sin{\phi}] \sin{i}\\
-(v_r \cos{\theta} - v_{\theta} \sin{\theta}) \cos{i} \nonumber
\end{eqnarray}
\noindent where $i$ is the inclination angle of the system.  The disk is 
`face-on' when $i=0^{\circ}$ and `edge-on' when $i=90^{\circ}$.         

The line emissivity $\epsilon(n,T)$ is calculated with the software package 
{\it fivelevel}, which was kindly provided by G. Mellema (private communication).
This is a key step since the lines we will focus on have markedly different 
critical densities.

Finally, we assume that our simulations have a typical solar abundance of 
elements.  We adopt the solar abundance ratios of Anders \& Grevesse 
(1989).  We also assume that all of the atoms are in the ionization state 
of interest.  For example, when calculating the $6731$\AA \, line, we assume 
100\% of the sulfur atoms in the disk wind are singly-ionized.

Line profiles are constructed by integrating eqn. (14)  over $r = 0.1 - 20 
r_g$, $\theta = 0^{\circ} - 90^{\circ}$, and $\phi = 0^{\circ} - 360^{\circ}$ 
(i.e., one hemisphere only, see discussion below) over a range of velocities. 
 Since the simulations are 2-dimensional, we assume azimuthal symmetry in these 
calculations.  Total integrated forbidden line luminosities and surface 
brightnesses are calculated by integrating eqn. (14) over all velocities.

\subsection{Global properties: ionized forbidden lines}

The majority of observations of T Tauri systems to date are unresolved (e.g. 
HEG95).  The nature of unresolved observations is such that they can be used 
to directly constrain only the {\it global} properties of these systems, such 
as the total forbidden line luminosity and the mean spectrum of the system. 
 Increasingly, however, high resolution instruments, such as the Space 
Telescope Imaging Spectrograph (STIS) on board the Hubble Space Telescope, are 
resolving nearby systems (e.g., Bacciotti et al. 2000, 2002).  In such cases, 
it is becoming possible to measure {\it radial} properties, such as forbidden 
line surface brightness profiles and spatially-resolved spectra.  Therefore, 
we are interested in examining both the global and `structural' observational 
properties of the present PDW model.  We examine here (and in \S 4.3) the 
observational consequences in the event the system is unresolved, while in \S 4.4 
we focus on the observational properties of a resolved system.

Plotted in Figure 8 are the predicted line profiles for the NII and SII lines 
as a function of inclination angle for 
{\epsscale{1.0}
\plotone{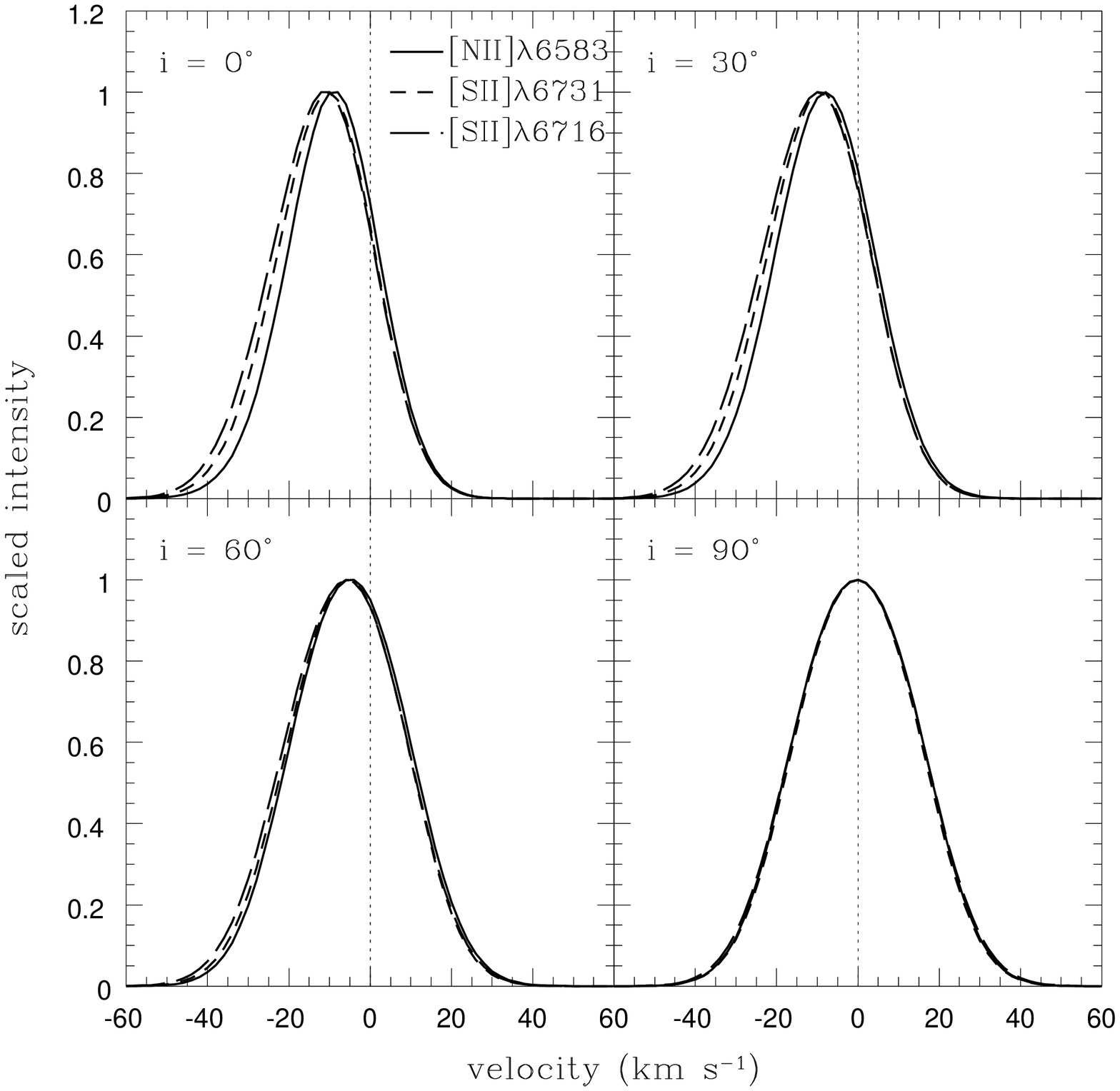}
{Fig. 8. \footnotesize
Predicted line profiles as a function of inclination angle for the PDW model.
The lines have all been arbitrarily normalized so that their peaks
are equal to 1.
}}
\vskip0.1in
\noindent
the PDW model.  The line profiles are 
calculated assuming $M_* = 1 M_{\odot}$ and $\Phi_* = 10^{41}$ s$^{-1}$.  The 
intensity of the lines have been scaled to make the peaks equal to unity in 
order to bring out differences in the shifts/shapes of the predicted profiles.

Focusing on the upper left-hand panel, we see that all three lines have been 
shifted to the blue by approximately $10$ km s$^{-1}$.  A blue-shift is 
expected since the disk is face-on, the outflow is moving away from the 
disk, and only one hemisphere was simulated.  Had we simulated both ionized 
hemispheres, but with no neutral disk in between, we would observe no shift 
whatsoever.  The presence of a thick neutral disk, however, acts as an 
occulting surface (with effectively infinite optical depth).  Since genuine 
observations also show blue-shifts similar to those plotted in Fig.\ 8 \ (e.g., 
HEG95), this implies that thick neutral disks are present in these systems.

The two sulfur lines are shifted more to the blue than the nitrogen line (by 
$\approx 2$ km s$^{-1}$).  The reason for this is that the sulfur lines have 
lower critical densities than the nitrogen line.  Thus, they are more sensitive 
to the outer, low-density regions of the atmosphere, where the gas is flowing 
at a higher velocity.  The widths (FWHM) of the three lines plotted in the 
upper left-hand panel range from $\approx 26-29$ km s$^{-1}$, with the sulfur 
lines being slightly wider than the nitrogen line.  This can also be attributed 
to the lower critical densities of the sulfur lines.

Examining the other three panels in Fig. 8, it is clear that as the inclination 
angle increases the line profiles become slightly wider and less shifted to 
the blue (see Table 1).  As the disk approaches the edge-on orientation the 
Keplerian nature of the disk is more observable and the radial outflow 
component is no longer solely flowing away from the observer.  At 
$i=90^{\circ}$, there are equal amounts of gas flowing towards and away from 
the observer, hence there is no shift.

{\epsscale{1.0}
\plotone{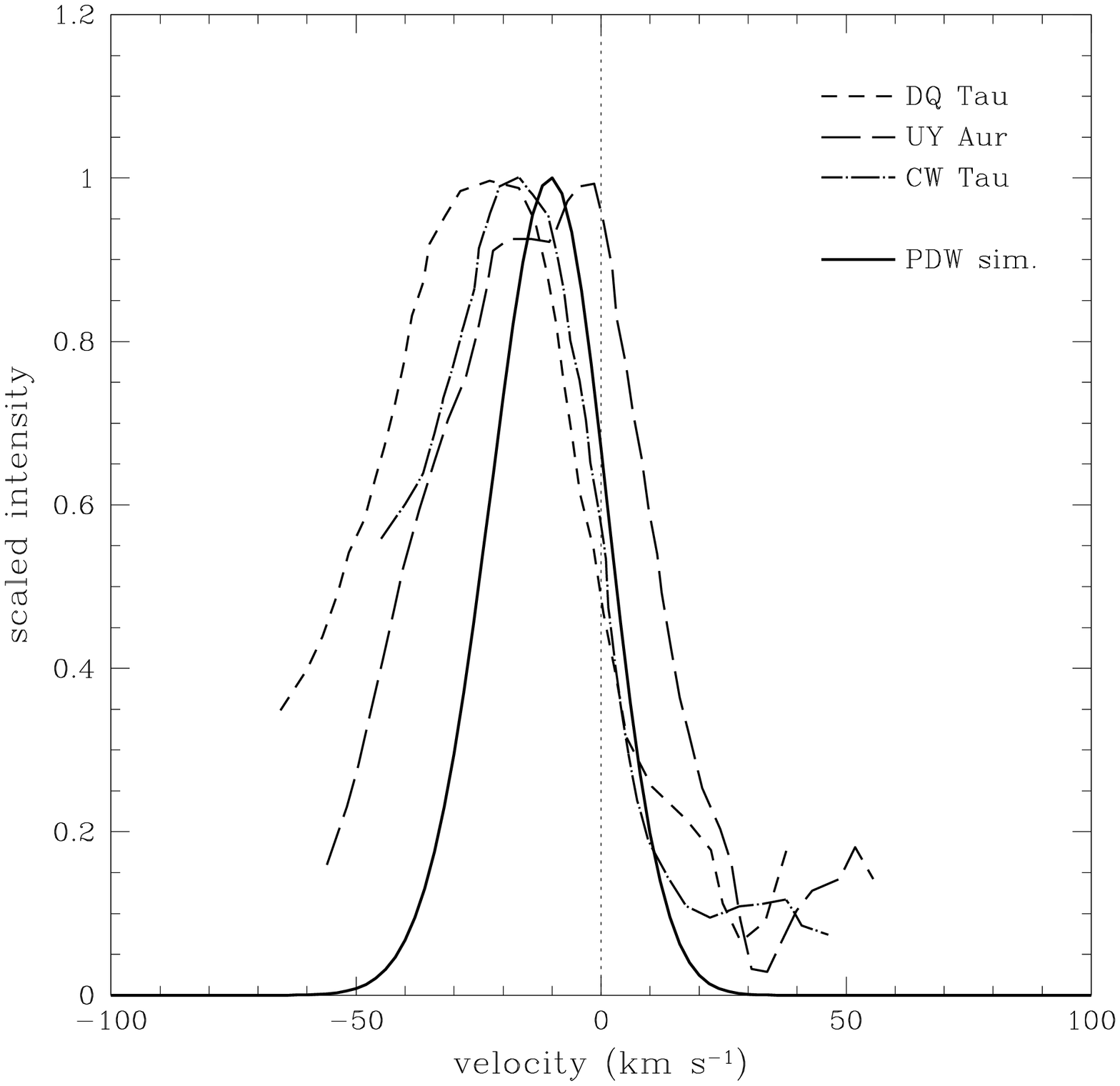}
{Fig. 9. \footnotesize
Comparison of [SII]$\lambda6731$ line profiles.  The PDW model profile is
calculated assuming $\Phi_* = 10^{41}$ s$^{-1}$, $M_* = 1 M_{\odot}$, and $i =
0^{\circ}$.
 }}
\vskip0.1in

How do the predicted line profiles 
compare with observed profiles?  HEG95 found that the forbidden line emission 
from T Tauri stars could be divided into two components, `low-velocity' and 
`high-velocity'. The high-velocity component correlates well with central jet 
diagnostics and shows extent in resolved observations. The low-velocity 
component, however, does not correlate with the central jet diagnostics and 
appears unresolved in the images. The low-velocity component appears to be 
associated with a disk wind (Kwan \& Tademaru 1988; 1995; Hirth, Mundt, \& Solf 
1997) similar to the winds examined in the present study.  Thus, we compare our 
results with the observed `low-velocity' component only. 
                                                                                      
Plotted in Figure 9 is a comparison of predicted and observed [SII]$\lambda6731$ 
line profiles.  The predicted line profile is calculated for a face-on PDW  
model and assumes $M_* = 1 M_{\odot}$ and $\Phi_* = 10^{41}$ s$^{-1}$. 
We note that changing these parameters (within reason) has only a small effect 
on the resulting line width and blue-shift.  The observed profiles were taken 
from Fig. 14 of HEG95.  All of the lines (both theoretical and observed) have 
been scaled such that their peaks are equal to 1.  Thus, the plot should be 
used to compare widths and blue-shifts only.  A comparison of integrated 
luminosities is given below.
                                                                                      
It is clear that the predicted profile is not a perfect match to the data.
The observed lines show a large amount of scatter in the blue-shift of their
peaks.  However, we find it encouraging that the predicted peak of our profile 
falls in between and quite close to the peaks of the observed profiles.  Our 
model predicts comparable, although slightly narrower, line widths 
than seen in the profiles of DQ Tau and CW Tau (UY Aur is clearly much wider and 
possibly contains two components).  Thus, we would argue that the 
thermally-driven PDW model provides a simple explanation for the kinematics of 
the observed flows.  It would be interesting to understand the slight difference 
in predicted and observed line widths, however.  As indicated by Table 1, an 
inclined system will give rise to increased line widths.  This probably cannot 
explain the difference entirely, however.  Below, we argue that a central 
gap in the disk may be necessary in order to explain the observed luminosities 
of the NII line at 6583\AA.  In terms of flux, a central gap would obviously 
give more weighting to the high velocity gas at large radii.  As explained in \S 
4.4, this, in turn, will give rise to increased line widths and could possibly 
explain this slight systematic difference.  Alternatively, a small boost from 
a central `X-wind' could also give rise to increased line widths.

Figure 10 presents the integrated line luminosities for the sulfur and 
nitrogen lines as a function of $M_*$ and $\Phi_*$.  Typically, the model 
predicts luminosities of a few times $10^{-7} L_{\odot}$ up to a few times 
$10^{-4} L_{\odot}$, with systems that have a higher central stellar mass being 
more luminous.  The trend of higher luminosities with higher central stellar 
masses (for a fixed value of $\Phi_*$) is easy to understand.  If the 
emissivity per ion, $\epsilon$, scales as $n$ (as would be the case if critical 
densities were unimportant), we should expect the luminosity to be independent 
of the mass since $L \propto \epsilon \ n \ dV \propto n^2 \ dr^3 \propto 
M_*^{-3} \ M_*^3$ (see eqns. 12-13).  However, the sulfur and nitrogen lines 
have critical densities that are comparable to the densities found in the 
outflow, thus $\epsilon$ does not quite scale as the density and, hence, $L 
\propto M_*^{\beta}$ where $\beta > 0$.  This is confirmed by comparing the NII 
relation with the SII relations.  Note that the NII line luminosity does not 
depend as strongly on $M_*$.  The reason for this is that NII has a higher 
critical density than does SII and, thus, a smaller fraction of outflow has 
densities equal to or exceeding the critical density.  It also explains the 
apparent bends in the $L$ vs. $M_*$ relations, since the critical densities 
become less relevant for systems with higher central stellar masses (eqn. 13).
The above arguments hold true for fixed values of $\Phi_*$, but it is worth 
noting that $\Phi_*$ will probably have some mass dependence (e.g., HJLS94).  
Thus, the two parameters are not entirely independent.

A comparison with the observational 
results of HEG95 is also presented in Figure 10.  For the data, we use the 
equivalent widths of the observed SII, NII, and OI forbidden lines (in Table 4 
of HEG95) to compute the observed integrated luminosities.  We note that these 
equivalent widths were derived from the low-velocity component only ($-60$ km 
s$^{-1} < u < 60$ km s$^{-1}$), which is appropriate for this comparison since 
we have not included jets in our hydrodynamic simulations.
                                                                                      
The top panels of Fig.\ 10 illustrate that the photoevaporative disk wind model 
predicts SII luminosities that are in excellent agreement with the observational 
data of HEG95.  This is remarkable given that the observed systems span more than 
an order of magnitude in $M_*$ and more than two orders of magnitude in $L$ and
that there was essentially no fine-tuning of the model parameters.
                                                                                      
The agreement with the observed NII luminosities (lower panel of Fig. 10),
however, is not as good.  In particular, our model tends to be
over-luminous with respect to the data.  Can the difference between the model
and data for the NII line be reconciled without spoiling the excellent
agreement between the observed and predicted SII luminosities?  As noted
before, the NII line has a higher critical 
{\epsscale{1.0}
\plotone{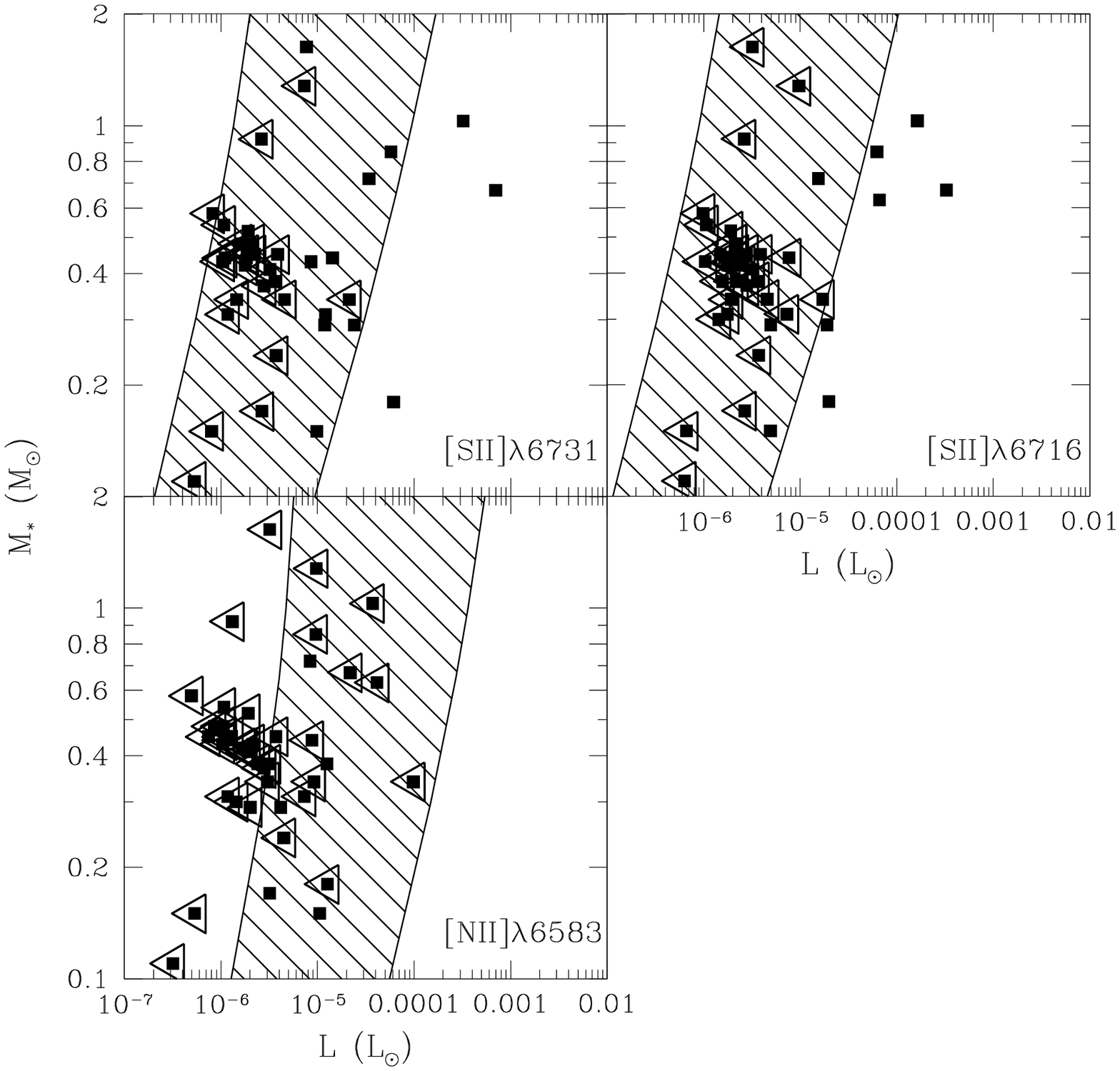}
{Fig. 10. \footnotesize
Predicted total line luminosities for the disk wind models.  The hashed swath
represents the predictions of the PDW model.
The low luminosity bound is set by assuming
$\Phi_* = 10^{40}$ s$^{-1}$ and $i = 0^{\circ}$, while the high luminosity
bound is set by assuming $\Phi_* = 10^{42}$ s$^{-1}$ and $i = 90^{\circ}$.
Solid squares are the observational data of HEG95, which is discussed in \S 5.
Open triangles pointing to low luminosities indicate upper limits.
}}
\vskip0.1in
\noindent
density than the SII
lines.  Thus, it is more sensitive to the central regions of the disk wind
than are the SII lines.  A potential way of reconciling both the NII
and SII lines simultaneously is if the central region of the ionized disk
atmosphere is either less dense than assumed by our PDW model
or entirely removed from the system.  HJLS94 considered the effect which strong
central winds might have on the structure of the ionized disk atmosphere and
concluded that although the inner disk atmosphere might be severely limited,
the outer evaporating disk would not be severely affected. Thus, it is
possible that the absence of a strong low-velocity NII component in the 
observations is evidence for the lack of an inner ionized atmosphere. 

\subsection{Global properties: neutral forbidden lines}

We have purposely left the discussion of OI forbidden emission lines until 
now.  This is because accounting for the observed line strength of 
[OI]$\lambda6300$ and [OI]$\lambda5577$ is difficult in the PDW model 
since most of the oxygen in the wind should be ionized.  Indeed, the 
similarity in the ionization potential of both hydrogen and oxygen, and the 
ensuing charge exchange between the two species, should ensure that the 
fraction of neutral oxygen is similar to the fraction of neutral hydrogen 
(Osterbrock 1989).  Thus, only at the location where the disk wind crosses the 
ionization front might one expect to find significant OI emission.  Given that 
the critical densities for these lines are even higher than for nitrogen, the 
line emission is dominated by the dense gas residing in the interior of the 
disk and near the base of the disk, just the location where the oxygen may be 
significantly neutral.  Therefore, it is not immediately clear that the present 
model is completely incapable of explaining the observed OI emission. 
                                                                                      
Indeed, externally evaporated disks are observed to produce strong OI emission 
(Bally et al. 1998) which is spatially confined to both the surface of the disk 
and the ionization front, with about equal total intensity in each component. 
The disk OI emission must, therefore, arise within the photodissociation region 
(PDR) of the flow, below the zone where the ionizing (EUV) photons penetrate.

St\"orzer \& Hollenbach (1998) provide an explanation for this strong OI 
emission from externally evaporated disk surfaces.  Theoretical calculations
of the dissociation of oxygen from OH (van Dishoeck \& Dalgarno 1983) reveal
that a significant fraction of the oxygen is left in an excited state.
St\"orzer \& Hollenbach determine the dissociation rate of OH and calculate
the resulting OI line luminosity.  The enhanced far ultraviolet radiation 
field, due to the proximity of a massive star, and the high number density 
within the PDR, produce a strong OI signature with an [OI]$\lambda6300$ surface 
brightness of $2.2 \times 10^{-2}\,$ ergs\,s$^{-2}$\,sr$^{-1}$. For typical disk 
sizes of 100\,AU this results in a total line luminosity of $\sim 
10^{-4}\,L_{\odot}$.  These values provide excellent agreement with the 
observations of Bally et al.\ (1998).  However, because the OI layer is 
confined to the base atmosphere, where the outflow velocities are smallest, 
the resulting lines are likely to have only modest blue-shifts.
                                                                                      
St\"orzer \& Hollenbach (2000) discuss the modification of the externally
evaporating disk model for the case of disks evaporated by their central
stars and determine that the far ultraviolet (FUV) radiation field is not strong
enough to provide significant OI line luminosities.  This result, however,
does not take into account the possibility of an ionized atmosphere
redirecting EUV photons toward the disk.  Indeed, the photoevaporating disk
model of SJH93 and HJLS94 requires the existence of a dense PDR sandwiched
between the ionized wind and the molecular disk.  

We check this idea quantitatively for the PDW model.  The EUV 
photons come from a 10,000 K source and most will have energies near the Lyman 
limit of 13.6 eV.  The cross section for hydrogen atoms to absorb these photons 
is of order $5 \times 10^{-18}$ cm$^2$.  Thus, the EUV photons penetrate 
a neutral column of $\approx 2 \times 10^{17}$ cm$^2$.  This means that the 
ionization 
front, the region where the gas is still 10,000 K but has a significant 
fraction of neutral oxygen and hydrogen, has a column density of H atoms of 
$\approx 2 \times 10^{17}$ cm$^2$.  We have determined the zone at the base of 
atmosphere in the PDW simulation which contains this column density for a 
range of stellar masses and ionizing photon rates.  Typically, we find that the 
predicted blue-shifts are of order $3-4$ km s$^{-1}$, which compares quite 
favorably to the observations of HEG95 (they find an average offset of $\sim 5$ 
km s$^{-1}$ for neutral oxygen lines), but the integrated luminosities are only 
of order $10^{-7}-10^{-6} L_{\odot}$, which is about two orders of magnitude 
lower than observed (HEG95).  Thus, without some kind of modification, the 
present model cannot easily accommodate the strength of the observed OI emission 
from T Tauri stars.  

\subsection{Radial properties}

A potentially powerful test of the model, which will be possible 
within the coming years, is through comparisons with observed structural 
properties.  Plotted in Figure 11 are predicted nitrogen and sulfur line 
profiles as a function 
{\epsscale{1.0}
\plotone{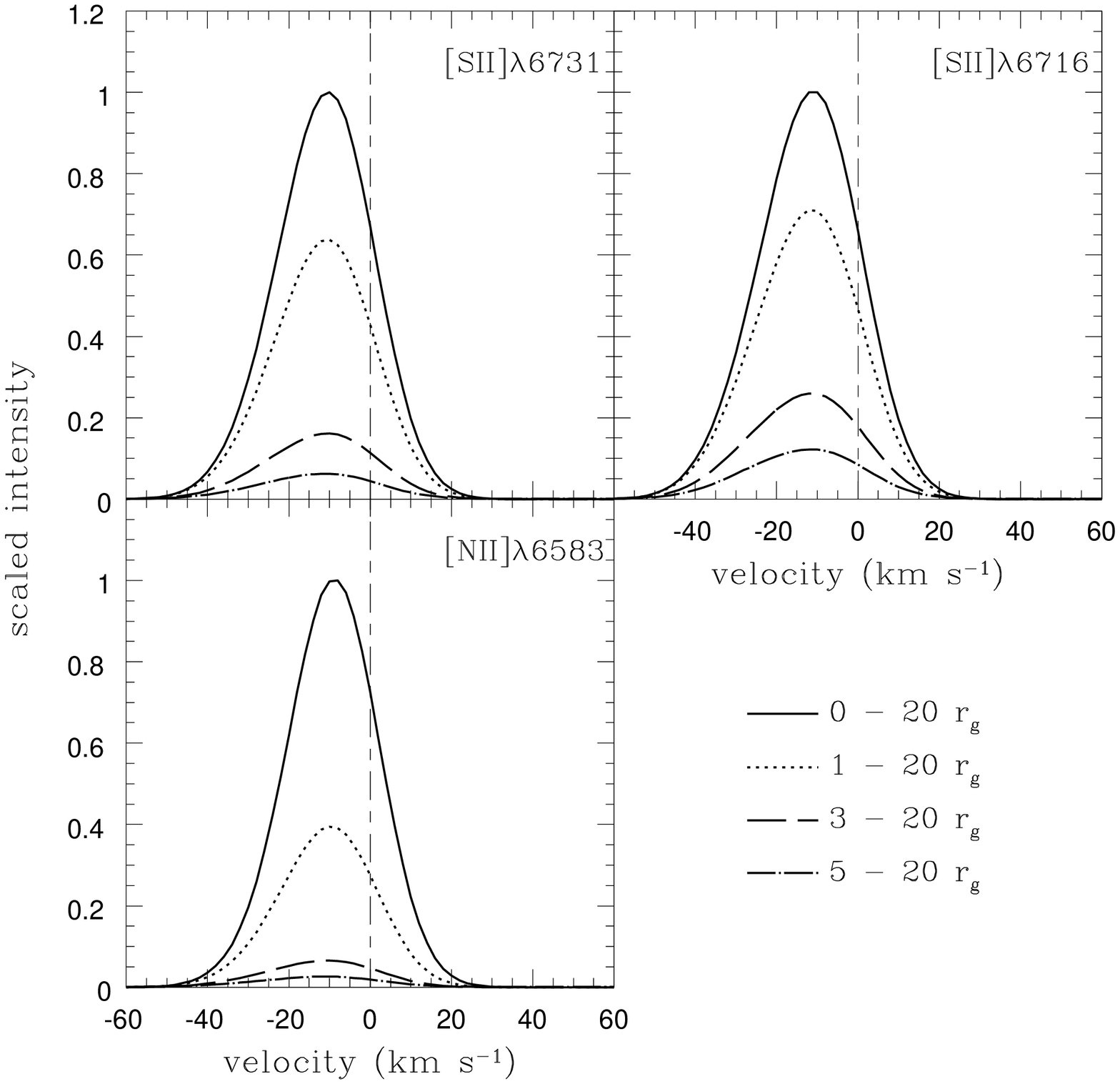}
{Fig. 11. \footnotesize
Predicted line profiles as a function of radius for the face-on ($i =
0^{\circ}$) PDW model.  The lines have all been normalized so that the
peak of the $0 - 20 r_g$ line (i.e., the entire disk) equals 1.
}}
\vskip0.1in
\noindent
of projected radius for the face-on PDW model.  The 
lines are calculated assuming $M_* = 1 M_{\odot}$ and $\Phi_* = 10^{41}$ 
s$^{-1}$.  Each of the lines have been normalized such that the peak of the 
line profile for the entire system, i.e., for $0 - 20 r_g$, is unity.  Thus, in 
addition to showing how the shapes of the line profiles change as a function of 
projected radius, the plots indicate how the various regions of the disk 
contribute to the mean spectrum of the system.

Focusing on the shapes of the profiles in the three panels, it is clear that as 
one goes further from the projected center of the outflow the profiles become 
more skewed.  In particular, at large projected radii there are `wings' evident 
on the blue sides of the profiles.  The development of line wings has a
physical explanation.  At small projected radii from the center, most of the 
flux from the system originates from the dense ionized disk.  Because the 
velocity of the gas flowing off the ionized disk is fairly constant as a 
function of radius, a nearly perfect Gaussian line profile is the result.  
Since most of the flux comes from the central regions of the outflow, this also 
explains why the mean spectrum of the entire system is almost exactly Gaussian.  
However, as one moves to larger projected radii the role of the atmosphere 
above the disk, in terms of flux, becomes more important.  The gas in the 
atmosphere is flowing faster than the gas near the ionized disk (e.g., Fig. 6a), 
hence, the Gaussian profile is skewed toward the blue.  The 
increased importance of the atmosphere at large projected radii also causes the 
peak of  the line profile to shift further to the blue and the FWHM of the line 
to increase.  We demonstrate this 
quantitatively by fitting Gaussians to the line profiles.  The results are 
presented in Table 1.  While the shift in the peak is modest ($\sim 2$ km 
s$^{-1}$) in going from the mean spectrum of the system to just focusing on 
the outer regions of the system, the increase in the line width is actually 
fairly substantial ($\gtrsim 15$\%) and potentially observable.   

{\epsscale{1.0}
\plotone{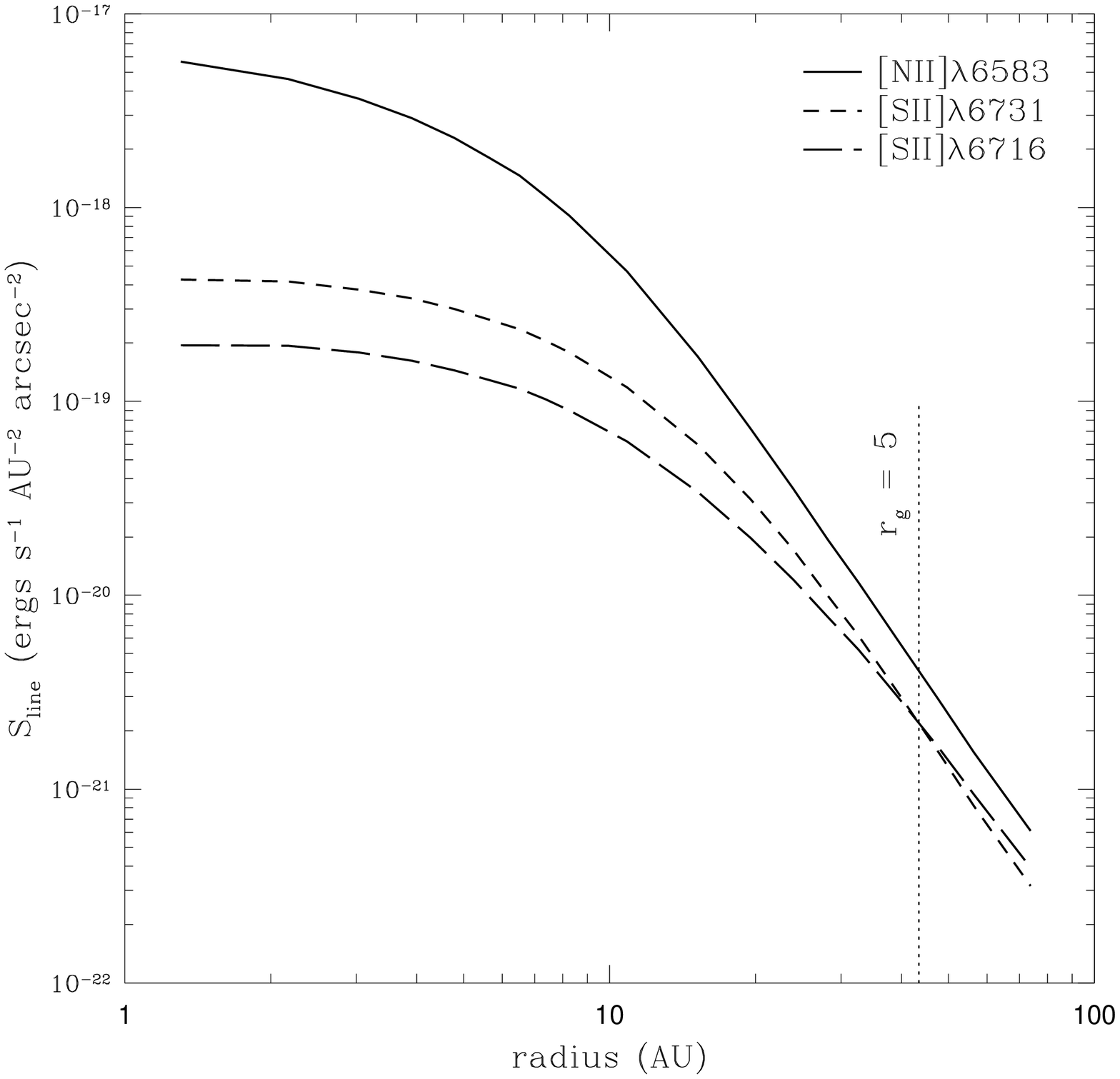}
{Fig. 12. \footnotesize
Predicted forbidden line surface brightness profiles for the face-on PDW model.
}}
\vskip0.1in

It is also evident in Fig. 11 that a larger fraction of the total flux from the 
nitrogen line comes from the central regions of the outflow than from the two 
sulfur lines.  The reason for this trend is that the nitrogen line is more 
sensitive to the high density regions of the atmosphere than are the sulfur 
lines.

Lastly, plotted in Figure 12 are the predicted surface brightness profiles for 
the sulfur and nitrogen lines.  These are also calculated assuming $M_* = 1 
M_{\odot}$ and $\Phi_* = 10^{41}$ s$^{-1}$.  The surface brightness profiles 
of the three lines all exhibit the same general features: a central core and 
a power-law drop off at large radii.  The size of the core is dictated by the 
critical densities of the lines, with the NII profile being the most 
concentrated because of the higher critical density of nitrogen.  At large 
radii, the profiles drop off as roughly $S_{line} \propto r^{-3}$.  This can be 
understood physically since $S_{line} \propto \epsilon \ n \ r \propto r^{-2} 
\ r^{-2} \ r$ ($\epsilon \propto n^{\sim 1}$ at large radii, low densities).  
The slight differences in the slopes of the three profiles are attributable to 
differences in the critical densities.

\section{Discussion and Conclusions}

The evidence for self-produced photoevaporative disk winds around low-mass stars 
is largely circumstantial.  The primary reason for this has been the
lack of detailed calculations on the outflow properties of these winds to 
compare with observations.  Photoevaporative disk wind models have, however, 
been shown to fit the observational data of externally-heated low-mass stars in 
the vicinity of massive stars (e.g., Johnstone et al. 1998; Bally et al. 1998; 
St\"orzer \& Hollenbach 1998) because the disk wind geometry and flow 
characteristics are much simpler.
                                                                                      
In this paper we use hydrodynamic simulations to compute the properties of the
photoevaporative disk wind model and, subsequently, the predicted observational
signatures.  We show that the hydrodynamic simulations converge to steady-state
solutions and that a number of general properties (e.g., scale height 
distribution, integrated disk wind mass-loss rate) agree well with the earlier
analytic results of SJH93 and HJLS94.  The outflows become nearly spherically
symmetric and match the analytic solution to the Parker solar wind problem at
large radii, as expected.  We further demonstrate that the photoevaporative
winds produce blue-shifted ionized forbidden lines with typical offset 
velocities of $\sim -10$ km s$^{-1}$, widths of $\sim 30$ km s$^{-1}$, and 
integrated luminosities of $10^{-7}-10^{-4} L_{\odot}$.  These values compare 
favorably with the observations of HEG95 (see Figs. 9 and 10).  The 
strength of the observed OI emission is difficult to account for with the 
present model, however.

The photoevaporative disk winds examined here are intuitive and 
physically-simplistic.  They rely on the availability of $\Phi_* = 10^{40-42}$
ionizing photons per second to be emitted by the central star disk accretion
shock.  Theoretical arguments support such high levels of ionizing emission
(e.g., Matsuyama et al. 2003a).  Observationally, it is extremely difficult to 
directly pin down the value of $\Phi_*$ for low-mass T Tauri stars.  HEG95 
measured the mass accretion rates onto the surface of the T Tauri stars, 
however, and these, in principle, can be used to estimate the value of $\Phi_*$.  
Given the mass accretion rate and the stellar radius, we calculate the accretion 
shock luminosity of the gas as it falls onto the star's surface via
                                                                                      
\begin{equation}
L_{\rm acc} = \frac{G M_* \dot{M}_{\rm acc}}{2 R_*}
\end{equation}
                                                                                      
\noindent where $\dot{M}_{acc}$ is the mass accretion rate and $R_*$ is the
radius of the T Tauri star (see also Matsuyama et al. 2003a).  Using the
inferred values of $L_{acc}$ and assuming black-body radiation with a 
characteristic temperature of 10,000 K (Johns-Krull et al. 2000; Gullbring et 
al. 2000), we indeed estimate $\Phi_* \sim 10^{40-42}$ s$^{-1}$.  This lends 
credence to the photoevaporative wind hypothesis.

Utilizing these ultraviolet photons, however, may be difficult. As mentioned
above, both the protostellar jet (Shang et al. 2002) and the accretion column
(Alexander et al. 2004) are likely to be optically thick to ionizing radiation,
protecting the disk.  Nevertheless, energy released within the accretion disk,
and in the fast-moving jet, may produce significant ionizing radiation.
Alternatively, the disk may be heated by another source such as FUV photons
(Johnstone et al. 1998, St\"orzer \& Hollenbach 1998).
                                                                                
The calculations presented here reveal that heating the disk surface to 
$10^4\,$K will produce a thermal wind with flow characteristics similar to those 
observed around T Tauri stars. Additionally, if the mass-loss rate is high 
enough, comparable to the evaporating disk model, the forbidden line profiles of
SII and NII can be explained. Curiously, if the $10^4\,$K wind is not ionized
by EUV photons the resultant OI forbidden line radiation also would be comparable
to that observed. Thus, the observations of HEG95 seem to require a partially 
ionized $10^4\,$K wind." 

Alternative models have also been proposed to explain the observed 
low-velocity component.  In fact, the prevailing view is that T Tauri winds are 
magnetically-driven (e.g., Anderson et al. 2003).  This is largely because 
strong magnetic fields seem to be the only way to power the collimated, high 
velocity jets that are responsible for the high-velocity component observed in T 
Tauri spectra.  In addition, magnetic wind models are able to reproduce a number 
of the observed features of the high-velocity component (e.g., Shang, Shu, \& 
Glassgold 1998; Garcia et al. 2001a, 2001b).  It is less clear, however, if the 
magnetic wind models can account for the low-velocity component, which was the 
subject of investigation in \S 4.  Below, we briefly discuss how the predicted 
low-velocity components of magnetic wind models compare with their 
observational counterparts.  We argue that the photoevaporative disk wind model 
proposed in the present study matches the current suite of observations at 
least as well magnetic wind models.
                                                                                      
Generally speaking, magnetic wind models may be divided into two classes: 
`X-winds' and `disk winds'.  The primary difference between these two classes is 
the physical extent of the magnetic field and, therefore, the location of the 
outflow launching radius.  In X-wind models, the magnetic field is contained 
entirely within the very inner regions of the disk immediately next to the star.  
Disk wind models, on the other hand, assume that there is a large scale magnetic 
field present which extends to several AU.  Unfortunately, it is extremely 
difficult to work out the exact observational predictions of such models in a 
self-consistent manner (Garcia et al. 2001a, 2001b).  This is one advantage the 
physically-simplistic photoevaporative disk wind model has over the magnetic wind 
models.
                                                                                      
While it is difficult to calculate the precise observational characteristics of
magnetic wind models, there have been a number of basic predictions made which
are testable with current observational data.  In terms of the X-wind models,
the predicted launching radius for the disk outflow is extremely close to the 
star.  However, recent high resolution observations suggest that the 
low-velocity component probably originates at `large' disk radii (e.g., 
Bacciotti et al. 2002; Anderson et al. 2003), suggesting that some other 
mechanism is responsible for the observed low-velocity component.  In fact, 
irrespective of the model, one should expect the outflow velocity to be of 
order the escape velocity at the region where the wind originated.  This is a 
strong argument that the low-velocity component must originate at moderate 
radii ($\sim 1-10$ AU).  
                                                                                      
Magnetic `disk wind' models, on the other hand, are able to produce both high 
and low-velocity components (e.g., Cabrit et al. 1999; Garcia et al. 2001a, 
2001b), with the low-velocity component being launched from reasonably large 
disk radii.  Furthermore, such models predict that the high velocity component 
is spatially extended while the low-velocity component is confined within the 
central (projected) regions of the system, which is in good agreement with 
observations.  However, the problem with `disk wind' models is that the level 
of ionization at large disk radii (i.e., at the launching radius) is too low to 
allow for strong magnetic coupling (e.g., Gammie 1996).  Hence, such models 
tend to under-predict the luminosity of the low-velocity component by a 
substantial margin (e.g., Garcia et al. 2001b).  As we have shown, 
however, the PDW model presented here (which contains no magnetic field) can 
naturally produce an outflow that matches a number of observed trends seen in 
low-mass T Tauri stars, so long as the photons can penetrate from the central 
star out to large radii (or if an external mechanism for producing EUV photons 
is present).   It is possible, however, that {\it both} photoevaporation and a 
large scale magnetic field could be contributing to the low-velocity component, 
especially since the higher level of ionization due to photoevaporation should 
increase the effectiveness of the magnetic disk wind.  This two component model
could possibly explain the relatively large amount of scatter in the forbidden 
line profiles and luminosities and why some stars apparently show multiple 
low-velocity components (e.g., UY Aur, see Fig. 9).

Additionally, it is likely that an interface between the fast jet-like 
wind arising from the inner disk will modify these results.  HJLS94 modeled the 
effect of a strong stellar wind on the ionized atmosphere and concluded that 
only the inner regions would be affected. This suggests that the results of the 
calculations in this paper would need to be modified only in the central densest 
regions, resulting in somewhat lower line fluxes. This would predominantly 
affect transitions having high critical densities such as the nitrogen line. 
Moreover, Matsuyama et al. (2003a) found that gaps could form in the inner 
part of the disk, near $r_g$, due to the erosive power of photoevaporation 
coupled to the viscous evolution of the disk.  Although the Matsuyama models 
should be recalculated, taking into account the results of this paper, these 
gaps would remove the inner densest part of the ionized atmosphere produced in 
this paper. An evolutionary sequence might be visible in the low-velocity 
spectra from T Tauri stars dependent upon the state of the underlying disk. 
Detailed calculations, taking into account each of these factors, should be 
undertaken.
                                                                                      
\vskip0.1in
\noindent The authors wish to thank the referee, David Hollenbach, for suggesting
significant improvements to the paper. Thanks also to G. Mellema for providing his 
software package for calculating forbidden line emissivities.  I.G.M. thanks A. Babul
for useful discussions.  A.F. acknowledges financial support from J.F. Navarro.
I.G.M. is supported by a postgraduate scholarship from the Natural Sciences and
Engineering Research Council of Canada (NSERC). The research of D.J.
has been supported by an NSERC Discovery Grant. D.R.B. thanks NSERC for 
financial support.

\begin{deluxetable}{ccccccc}
\tablecaption{Line properties
\label{tab1}}
\tablewidth{30pc}
\tablenotetext{~}{}
\startdata
\tableline
\tableline
        & [NII] & [NII] & [SII] & [SII] & [SII] & [SII]\\
        & 6583\AA & 6583\AA & 6731\AA & 6731\AA & 6716\AA & 6716\AA\\
\tableline \\
        & FWHM & shift & FWHM & shift & FWHM & shift\\
$i$ (deg.)  & (km s$^{-1}$) & (km s$^{-1}$) & (km s$^{-1}$) & (km s$^{-1}$)
& (km s$^{-1}$) & (km s$^{-1}$)\\
\tableline
\\
0 & 26.5 & -9.2 & 28.2 & -10.9 & 29.4 & -11.8\\
30 & 28.7 & -8.1 & 29.8 & -9.6 & 31.0 & -10.3\\
60 & 33.2 & -4.9 & 33.2 & -5.6 & 34.2 & -6.0\\
90 & 35.4 & 0.0 & 34.9 & 0.0 & 35.8 & 0.0\\
\\
\tableline
\\
        & FWHM & shift & FWHM & shift & FWHM & shift\\
Region  & (km s$^{-1}$) & (km s$^{-1}$) & (km s$^{-1}$) & (km s$^{-1}$)
& (km s$^{-1}$) & (km s$^{-1}$)\\
\tableline
\\
0 - 20 $r_g$ & 26.5 &-9.2 & 28.2 & -10.9 & 29.4 &-11.8\\
1 - 20 $r_g$ & 28.0 &-10.3 & 29.2 & -11.4 & 30.4 & -12.3\\
3 - 20 $r_g$ & 31.6 &-11.8 & 31.5 & -11.8 & 32.5 &-12.7\\
5 - 20 $r_g$ & 32.8 &-12.6 & 32.7 & -12.5 & 33.1 &-13.0\\
\\
\enddata
\end{deluxetable}

\end{document}